\documentclass[%
 reprint,
 twocolumn,
 amsmath,amssymb,
 aps,
]{revtex4}

\usepackage[utf8]{inputenc}
\usepackage{bm}
\usepackage{xcolor,soul}
\usepackage{chemformula}

\usepackage{braket}
\usepackage{blkarray}
\usepackage{amsthm}
\usepackage{multirow}
\usepackage{adjustbox}
\usepackage[colorlinks,linkcolor=blue,anchorcolor=blue,citecolor=blue,urlcolor=blue]{hyperref}

\usepackage{cleveref}
\Crefname{equation}{Eq.}{Eqs.}
\Crefname{figure}{Fig.}{Figs.}
\Crefname{tabular}{Tab.}{Tabs.}
\usepackage{comment}

\newcommand{\prepare}{\mbox{P{\scriptsize REPARE}}}

\newcommand{\select}{\mbox{S\scriptsize ELECT}}

\theoremstyle{definition}

\theoremstyle{definition}

\theoremstyle{remark}

\begin{document}
\title{Beyond asymptotic reasoning: { the practicalities of a quantum }ground state projector based on the wall-Chebyshev expansion}
\author{Maria-Andreea Filip}
\email{maf63@cam.ac.uk}
\affiliation{Yusuf Hamied Department of Chemistry, University of Cambridge, Lensfield Road, Cambridge CB2 1EW, United Kingdom}
\author{Nathan Fitzpatrick}
 \email{nathan.fitzpatrick@quantinuum.com}
\affiliation{Quantinuum, 13-15 Hills Road, CB2 1NL, Cambridge, United Kingdom}
\begin{abstract}
   { We consider a quantum algorithm for ground-state preparation based on a Chebyshev series approximation to the wall function. In a classical setting, this approach is appealing as it guarantees rapid convergence. We analyze the asymptotic scaling  and success probabilities of different quantum implementations and provide numerical benchmarks,  comparing the performance of the wall-Chebyshev projectors with current state-of-the-art { approaches}. We find that this approach requires fewer serial applications of the Hamiltonian oracle to achieve a given ground state fidelity, but is severely limited by exponentially decaying success probability. However, we find that some implementations maintain non-trivial success probability in regimes where wall-Chebyshev projection leads to a fidelity improvement over other approaches. As the wall-Chebyshev projector is highly robust to loose known upper bounds on the true ground state energy, it offers a potential resource trade-off, particulary in the early fault-tolerant regime of quantum computation.}
\end{abstract}

\maketitle

\section{Introduction}
Imaginary time evolution (ITE) is an effective means to obtain the ground state of quantum systems. The formalism is obtained by a Wick rotation\cite{Wick1954} of the time-dependent Schr\"odinger equation
\begin{equation}
    i\frac{\partial \Psi(t)}{\partial t} = \hat H \Psi(t),
\end{equation}
by transforming $it \rightarrow \tau$ to give the so-called imaginary time Schr\"odinger equation
\begin{equation}
    \frac{\partial \Psi(\tau)}{\partial \tau} = -\hat H \Psi(\tau),
    \label{eq:sch}
\end{equation}
where $\hat H$ is the problem Hamiltonian, $\Psi$ is the system wavefunction and $\tau$ is imaginary time. \Cref{eq:sch} has a solution of the form
\begin{equation}
    \Psi(\tau) = C e^{-\hat H \tau}\Psi(0),
\end{equation}
where C is some constant of integration. Given a set $\{\Psi_i\}$ of eigenfunctions of the Hamiltonian, with corresponding eigenvalues $\{E_i\}$, the wavefunction may be decomposed into
\begin{equation}
    \Psi(\tau) = C \sum_i c_i e^{-E_i \tau}\Psi_i,
    \label{eq:ci}
\end{equation}
where $\Psi(0) = \sum_i c_i \Psi_i$. Therefore, as $\tau \rightarrow \infty$, the right-hand-side of \Cref{eq:ci} will be dominated by the wavefunction with the lowest $E_i$,
so
\begin{equation}
    \lim_{\tau \rightarrow \infty} \Psi(\tau) \propto \Psi_0.
\end{equation}
Imaginary time evolution therefore provides a valuable alternative to variational wavefunction optimization. In the full Hilbert space, { for systems with non-degenerate ground states,}there is only one minimum of energy\cite{Burton2022}. {Using} restricted wavefunction ans\"atze leads to the appearance of local minima, generating very complex energy landscapes\cite{Choy2025}. This can cause conventional optimization algorithms to become stuck, while imaginary time evolution is guaranteed to converge to the global minimum of the ansatz, provided a sufficiently long time interval. Additionally, in the context of quantum computing, it allows for ansatz-free wavefunction optimization, allowing the overall Hilbert space ground state to be found. However, unlike the real time propagator $e^{-i\hat Ht}$, the imaginary time propagator $e^{-\hat H\tau}$ is nonunitary, making it challenging to implement on a quantum computer. Various approaches have been developed to address this problem. 

Variational imaginary time evolution\cite{Jones2019, McArdle2019, Benedetti2021} (VITE) uses an ansatz to approximate the imaginary time-evolved wavefunction and then optimizes it according to some variational principle\cite{Yuan2019}. This is a hybrid quantum-classical algorithm and therefore appropriate for noisy intermediate-scale quantum (NISQ) devices. However, the original implementation of VITE\cite{Jones2019, McArdle2019} requires matrix inversions to obtain the ansatz parameters and is therefore highly susceptible to noise. Additionally, like most variational quantum algorithms, VITE may suffer from optimization problems, such as barren plateaus\cite{McClean2018}, which preclude convergence.

In quantum imaginary time evolution (QITE)\cite{Motta2020, Nishi2021, Sun2021}, the renormalized action of the imaginary time propagator is mapped onto a unitary operator, which can be determined from multiple sets of linear equations. The storage and measurement required scales as the exponential of the correlation length, which increases during the course of the time evolution. This scaling can be improved by imposing additional locality constraints\cite{Motta2020}. Alternatively, circuit depth can be reduced by lifting the locality constraint\cite{Nishi2021}.

Finally, probabilistic imaginary time evolution (PITE)\cite{Liu2021,Kosugi2022,Turro2022,Silva2023,Leadbeater2024} uses a block-encoded form of the imaginary time evolution operator. This is often done by encoding a short-time propagation operator and applying it repeatedly. While this decreases the error in the encoding of the operator itself, the probability of success decays exponentially with the number of repetitions. In addition to reasonably deep circuits and ancilla qubit requirements, which scale logarithmically with the number of terms in the Hamiltonian for linear combination of unitaries (LCU) based encoding approaches, this makes PITE methods more suitable for fault-tolerant quantum computers than current NISQ devices.

Encodings which do away with the need for repeated application of the unitary as well as the large number of ancillas have also been developed. One such approach uses the Quantum Eigenvalue Transform with Unitary block-encoding\cite{Dong2022} (QETU) of a polynomial approximation to the imaginary time evolution operator\cite{Chan2023}.

Alternative polynomial expansions can also be used to project onto the ground state of a quantum system of interest. In conventional quantum chemistry, particularly in the area of Monte Carlo simulations\cite{Anderson1976,Zhang2003,booth_fermion_2009,thom_stochastic_2010}, it is common to employ the first-orderer Taylor expansion of the imaginary time-evolution operator,
\begin{equation}
    e^{-\hat H \delta\tau} \approx 1 - \delta \tau\hat H ,
\end{equation}
as a projector, with the ground state obtained in the infinite-time limit as
\begin{equation}
    \Psi_0 = \lim_{n\rightarrow\infty} (1 - \delta \tau\hat H)^n\Psi(0).
\end{equation}
In a quantum setting, the hybrid Projective Quantum Eigensolver (PQE) method is based on this projection approach\cite{Stair2021}.

The original QETU\cite{Dong2022} approach employed a polynomial approximation of a shifted step function
\begin{equation}
    {\text{step}}(x)=
    \begin{cases}
        1, x\leq \mu\\
        0, x>\mu
    \end{cases}
\end{equation}
for ground state preparation. In the absence of a good estimate of the ground state energy, this may be employed as a projector onto a low-energy subspace, which can be combined with binary search approaches to find the ground state\cite{Lin2020b}. A more general approach for arbitrary eigenvalue filtering uses a quantum signal processing (QSP) implementation of 
\begin{equation}
    R_l(x; \Delta) = \frac{T_l(-1+2\frac{x^2-\Delta^2}{1-\Delta^2})}{T_l(-1+2\frac{-\Delta^2}{1-\Delta^2})},
    \label{eq:R_l}
\end{equation}
where $T_l(x)$ is the $l$-th Chebyshev polynomial of the first kind, defined as
\begin{equation}
   T_l(\cos\theta) = \cos(l\theta),
\end{equation}
and $\Delta$ is the spectral gap of the intended eigenvalue\cite{Lin2020a}.
Chebyshev polynomials have also been used in conventional quantum chemistry to implement a projector based on the wall function\cite{zhang_2016,zhao2024rapidly},
\begin{equation}
    {\text{wall}}(x)=
    \begin{cases}
        \infty,\ x\leq-1\\
        1,\ x=-1\\
        0,\ x>-1.
    \end{cases}
    \label{eq:wall}
\end{equation}
This can be considered as the limit as $\tau$ tends to infinity of the exponential propagator
\begin{equation}
    \lim_{\tau\rightarrow\infty}e^{-\tau (x-S)}=\text{wall}(x-S),
\end{equation}
and we can see that it must recover the ground state wavefunction.

In this work, we introduce a quantum algorithm for ground state preparation based on the wall-Chebyshev projector, and benchmark its practical performance against leading approaches such as quantum imaginary time evolution, step functions and eigenvalue filtering. { We consider query complexity, success probability and numerical performance, particularly in scenarios where accurate ground-state energy estimates are unavailable. It is shown that in these regimes step function and eigenstate filtering projector techniques fail to reach the ground state. The wall-Chebyshev operator can successfully identify the ground state in this regime, albeit with small success probabilities.} 

\section{The wall-Chebyshev projector}
\label{sec:cheby}
\begin{figure}
    \centering
    \includegraphics[width=\linewidth, trim = 0 0 1.2cm 1.2cm, clip]{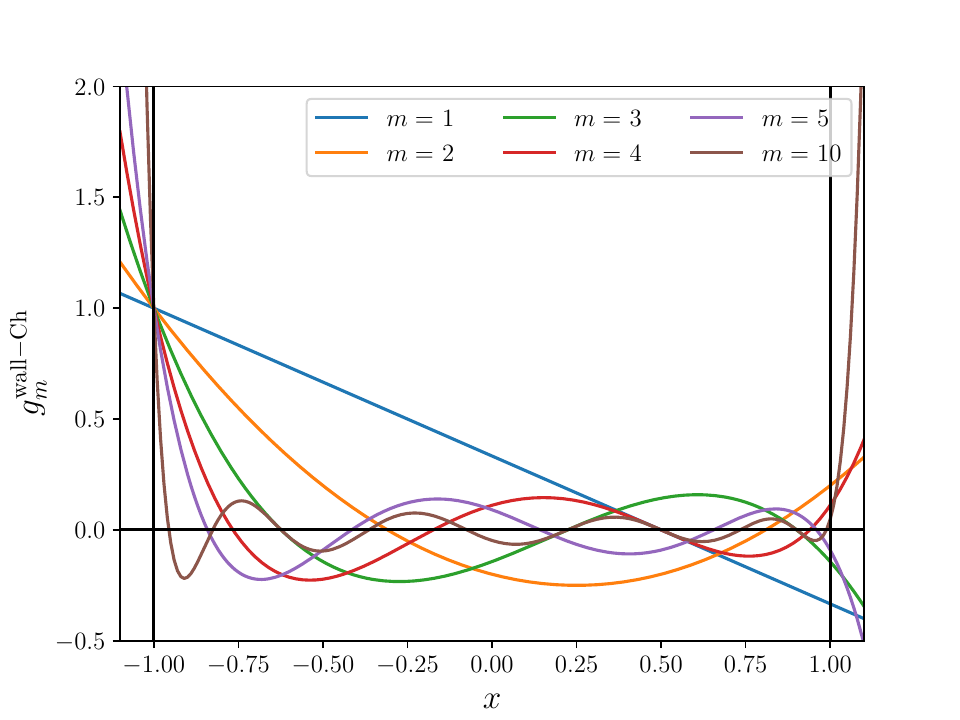}
    \caption{Different order Chebyshev polynomial expansions of the wall function.}
    \label{fig:cheby_poly}
\end{figure}

Chebyshev polynomials are naturally defined on the interval $[-1,1]$. An $m$-th order Chebyshev expansion of the wall function (\Cref{eq:wall}) on this interval is given by
\begin{equation}
\begin{split}
    G^{\text{wall-Ch}}_m(x)&= \frac{1}{1+2m}\sum_{k=0}^m (2-\delta_{k0}) (-1)^kT_k(x)\\
    &=\frac{1}{1+2m}\sum_{k=0}^m(2-\delta_{k0})T_k(-x),
\end{split}
\label{eq:big_G}
\end{equation}
where, $\delta_{ij}$ is the Kronecker delta. Some examples are shown in \Cref{fig:cheby_poly}. We can map the Hamiltonian eigenspectrum onto the $[-1,1]$ interval through the transformation
\begin{equation}
    \widetilde H = 2\frac{\hat H-E_0}{R} - 1,
    \label{eq:rescale}
\end{equation}
where $R = E_\mathrm{max} - E_0$ is the spectral range of the Hamiltonian, giving
\begin{equation}
    g^{\text{wall-Ch}}_m(\hat H)= \frac{1}{1+2m}\sum_{k=0}^m(2-\delta_{k0})T_k\big(1-2\frac{\hat H-E_0}{R}\big),
    \label{eq:small_G}
\end{equation}

This polynomial decomposition has a series of nodes in the interval $[E_0, E_0 + R]$ at
\begin{equation}
    a_{\nu}=E_0+\frac{R}{2}\left(1-\cos\frac{\nu\pi}{m+1/2} \right).
\end{equation}
See, for example, Appendix B in Ref. \cite{zhao2024rapidly} for a proof. We can then write:
\begin{equation}
    g^{\text{wall-Ch}}_m(\hat H) = \prod_{\nu=1}^m \frac{\hat H-a_{\nu}}{E_0-a_{\nu}}.
    \label{eq:wall-terms}
\end{equation}
As this is not an exact representation of the wall function, further increased overlap with the ground state may be obtained by repeating the application of this projector.
We define
\begin{equation}
    |\Psi^{(n,0)}\rangle=[g^{\text{wall-Ch}}(\hat{H})]^n|\Phi\rangle
\end{equation}
as the wavefunction after an integer number of applications of the projector, { where $\ket{\Psi}$ is an initial reference wavefunction. We} additionally define intermediate wavefunctions as
\begin{equation}
\label{eq:cheby_interm}
|\Psi^{(n,\mu)}\rangle = \left[\prod_{\nu=0}^{\mu}\frac{\hat{H}-a_{\nu}}{S-a_{\nu}}\right] [g^{\text{wall-Ch}}(\hat{H})]^n|\Phi\rangle .    
\end{equation}
where $S$ is an estimate of the unknown $E_0$.

In principle, for a given number $N$ of applications of the Hamiltonian, applying the order$-N$ wall-Chebyshev projector will give a better ground-state approximation than applying a lower order$-m$ projector $N/m$ times. However, lower-order projectors are generally more numerically stable, which may lead to better overall performance\cite{zhao2024rapidly}. Additionally, in a classical setting, repeated projection allows recomputation of $S$ between the application of different projectors, which can be used to assess convergence and increase the accuracy of the projector.

Implementing this projector requires estimates for the ground and highest excited states of the Hamiltonian. The latter can be estimated using an approximation to the Gershgorin circle theorem\cite{gersgorin_uber_1931} (see \Cref{app:gersh}) as
\begin{equation}
    \widetilde{E}_{N-1}=H_{N-1,N-1}+{\sum_{{j \neq N-1}}}|H_{N-1,{j}}|.
\end{equation}
{ where $N$ is the dimension of the Hilbert space of the problem.}
Using the same estimator for the ground state energy $S$, we define the estimated spectral range of the Hamiltonian as
\begin{equation}
    R ={ \beta} \widetilde{E}_{N-1} - S
\end{equation}
with ${ \beta} \geq 1$ to ensure the true spectral range of the Hamiltonian is less than $R$. This is important as, unlike the exact wall function, the Chebyshev approximation diverges not only for $x<-1$, but also for $x>1$.
The ground-state energy estimator may be initialised to the energy of the reference wavefunction $\ket{\Phi}$ and, if repeated applications of the projector are needed, it can be updated over the course of the propagation using the intermediate wavefunction $\ket{\Psi^{(n,0)}}$. For this purpose, one would need to estimate the expectation value of the energy $\langle \Psi^{(n,0)}|\hat H|\Psi^{(n,0)}\rangle$, which would require repetitions of the projection circuit. In practice, we find that, provided the original estimate of the ground state energy is higher than the true value, recomputing it is generally unnecessary.

\section{Complexity Analysis }

\begin{figure*}[!htbp]
    \centering
    
    \includegraphics[width = \textwidth, trim = 2cm 18.5cm 2 1.5cm, clip]{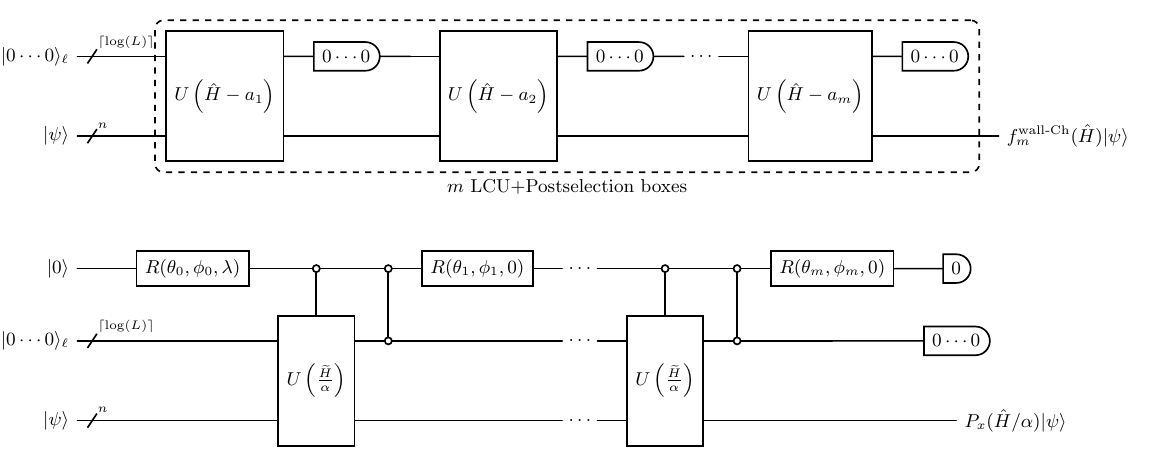}
    \caption{Encoding circuits for wall-Chebyshev projector of order $m$. \textbf{Top:} Product of LCUs corresponding to each term in Eq. 34. \textbf{Bottom:} GQSVT implementation following Ref. \cite{Sunderhauf2023}. $P_x$ can be either $P_1$ or $P_2$ as defined in Sec. IV, depending on the phases acting on the first ancilla qubit.}
    \label{fig:lcu_projector}
    \end{figure*}

The fidelity of the wavefunction obtained after applying a wall-Chebyshev projector of order $m$, labelled $g$ in the following, is given by

\begin{equation}
\begin{split}
    f&= \frac{\braket{\Psi_0|\Psi}}{\sqrt{\braket{\Psi|\Psi}}} = \frac{c_0g(E_0)}{\sqrt{\sum_i{|c_i|^2 \left|g(E_i)\right|^2}}}\\
    &= \frac{1}{\sqrt{1+\sum_{i>0}\left|\frac{c_i}{c_0}\right|^2\left|\frac{g(E_i)}{g(E_0)}\right|^2}}\\
    &\geq \frac{1}{1+\sqrt{\sum_{i>0}\left|\frac{c_i}{c_0}\right|^2\left|\frac{g(E_i)}{g(E_0)}\right|^2}}\\
    &\geq \frac{1}{1+\frac{1}{|c_0|}\max_{i>0}\left(\left|\frac{g(E_i)}{g(E_0)}\right|\right)\sqrt{\sum_{i>0}\left|c_i\right|^2}}\\
    &=\frac{1}{1+\frac{1}{|c_0|}\max_{i>0}\left(\left|\frac{g(E_i)}{g(E_0)}\right|\right)\sqrt{1-\left|c_0\right|^2}} \\
    & \geq \frac{1}{1+\frac{1}{|c_0|}\max_{i>0}\left(\left|\frac{g(E_i)}{g(E_0)}\right|\right)}.
\end{split}
\end{equation}

We can assume without loss of generality that $c_0$ is real and positive. In the limit of large $m$, { for sufficiently large values of $\beta$, such that the estimated $R$ exceeds the spectral range of the Hamiltonian}, $g$ decays with $E$, so $\max_{i>0}\left(\left|\frac{g(E_i)}{g(E_0)}\right|\right) = \left|\frac{g(E_1)}{g(E_0)}\right|$. Therefore,

\begin{equation}
    f \geq \frac{1}{1+\frac{1}{c_0}\left|\frac{g(E_1)}{g(E_0)}\right|}.
\end{equation}
$\left|\frac{g(E_{1})}{g(E_0)}\right|$ can be approximated as
\begin{equation}
    \left|\frac{g(E_{1})}{g(E_0)}\right|\approx \left|1+(E_1-E_0)g'(E_0) \right|\equiv\left|1-\Delta\gamma \right|\approx \exp(-\Delta \gamma), 
\end{equation}
where $\gamma=-g'(E_0) $ is called the convergence factor of $g$\cite{zhang_2016}. For the wall-Chebyshev projector, we derive $\gamma$ in Appendix \ref{app:gamma} and find it to be
\begin{equation}
    \gamma = \frac{2m(m+1)}{3R}.
\label{eq:gamma}
\end{equation}
Therefore
\begin{equation}
    \begin{split}
        \epsilon = 1 - f & \leq 1 - \frac{1}{1 + \frac{\exp(-\Delta \gamma)}{c_0}}\\
        &\leq \frac{\exp\left(-\frac{2\Delta m (m+1)}{3R}\right)}{c_0}
    \end{split}
\end{equation}
and the required polynomial order for a given error $\epsilon$ scales as
\begin{equation}
    m = \mathcal{O}\left(\Delta^{-1/2} (\log(\epsilon^{-1}c_0^{-1}))^{1/2}\right),
\end{equation}
which represents a polynomial improvement in asymptotic scaling relative to comparable projectors detailed in Appendix \ref{app:complexity}. 

While the scaling of $m$ with system size is very favourable, we note that the alternative approach of applying a fixed-order projector repeatedly also scales competitively. The error in the fidelity after applying the projector $g$ $n$ times is given by
\begin{equation}
\begin{split}
    \epsilon^{(n)} \leq { 1 -} \frac{1}{1+\frac{1}{c_0}\left|\frac{g(E_1)}{g(E_0)}\right|^n},
\end{split}
\end{equation}
so{,  by a similar argument to above,} the number of applications of the Hamiltonian scales as
\begin{equation}
    n = \mathcal{O}\left(\Delta^{-1}\log (\epsilon^{-1} c_0^{-1})\right),
\end{equation}
which is on par with state-of-the-art projectors, { as shown in \Cref{tab:asymptotics}. { All properties of $g$ required for this scaling also hold if we only have access to an upper-bound on $E_0$.} We note that this analysis does not account for the success probability of these approaches. This is a severe limitation in the wall-Chebyshev projector, as derived in the following section. The success probabilities of the other projectors considered here are discussed in  \Cref{app:complexity} and in general are lower bounded by $c_0^2$, if the ground state energy is known to high accuracy.}

\begin{table}[h]
    \centering
    \begin{ruledtabular}
    \begin{tabular}{ll}
        Method &  Projector query depth\\
        \hline
        \textbf{wall-Chebyshev} &   $\mathcal{O}\left(\Delta^{-1/2}\left(\log (\epsilon^{-1} c_0^{-1})\right)^{1/2}\right)$\\
        \textbf{wall-Chebyshev, iterative} &   $\mathcal{O}\left(\Delta^{-1}\left(\log (\epsilon^{-1} c_0^{-1})\right)\right)$\\
        Imaginary time evolution\cite{Silva2023} & $\mathcal{O}\left(\Delta^{-1}\log (\epsilon^{-1} c_0^{-1})\right)$\\
        Eigenstate filter\cite{Lin2020a} & $\mathcal{O}\left(\Delta^{-1}\log (\epsilon^{-1} c_0^{-1})\right)$\\
        Step function\cite{Lin2020b} & $\mathcal{O}\left(\Delta^{-1}\log (\epsilon^{-1}c_0^{-1})\right)$\\
    \end{tabular}
    \end{ruledtabular}
    \caption{Asymptotic scaling of the query complexity without success probability guarantees with respect to error for the wall-Chebyshev projector described in this work and alternative ground-state projection algorithms. Here, $\Delta$ is the gap between the ground and first excited states of the Hamiltonian, $c_0$ is the initial overlap with the ground state and $\epsilon$ is the target error. For comparison, we list the asymptotic scaling of comparable projection algorithms. Brief derivations of these scalings, following Refs. \cite{Lin2020a,Lin2020b,Silva2023}, are given in \Cref{app:complexity}.}
    \label{tab:asymptotics}
\end{table}

\section{Implementation and Success Probability}

\begin{figure}
    \centering
    \includegraphics[width=\linewidth]{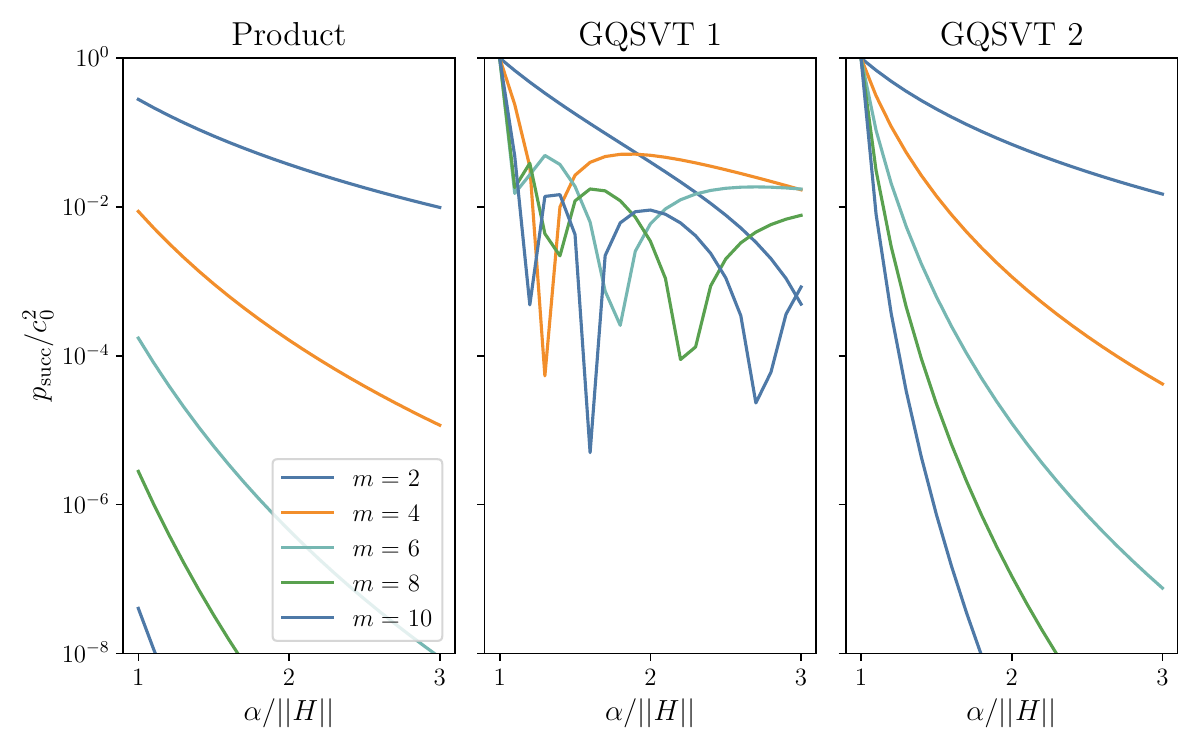}
    \caption{Theoretical success probability of different implementations of the wall-Chebyshev projector. The product form assumes the upper bound on $\alpha_v = \alpha + |a_v|$.}
    \label{fig:success_prob_alpha}
\end{figure}

For an $s$-qubit operator $A$, a unitary matrix $U$ acting on $s+b$ qubits corresponds to a $(\alpha, b, \varepsilon)$-block encoding of $A$ if
\begin{equation}
\| A - \alpha ( \langle \Pi |^b \otimes I ) U ( |\Pi\rangle^b \otimes I ) \| \leq \varepsilon
\end{equation}
For $\varepsilon = 0$, this corresponds to
\begin{equation}
    A = \alpha ( \langle \Pi |^b \otimes I ) U ( |\Pi\rangle^b \otimes I )
\end{equation}
and $U$ has the form
\begin{equation}
U =\ \begin{blockarray}{ccc}
    |\Pi\rangle|\psi\rangle & |\Pi_\perp\rangle|\psi\rangle &  \\
     \begin{block}{(cc)c}
     A/\alpha & * & \langle \Pi | \langle \psi|  \\
     * &  * & \langle \Pi_\perp | \langle \psi| \\
     \end{block}
     \end{blockarray}.
\end{equation}
$|\Pi\rangle$ is typically the $|0\rangle^{\otimes b}$ state and $|\psi\rangle$ is the wavefunction of the system register. The sub-normalisation $\alpha$ must be such that $||A||/\alpha \leq 1$, where $||\cdot||$ denotes the spectral norm of an operator. The success probability of applying a block-encoded operator is then given by 
\begin{equation}
    P = \frac{\braket{\psi|\hat A^\dagger A|\psi}}{\alpha^2}.
\end{equation}
A variety of options are available to obtain block-encodings of operators. Here we use the linear combination of unitaries (LCU) approach~\cite{Childs2012}, detailed in \cref{app:lcu}, to encode the Hamiltonian. { We then consider three alternative block encodings of the wall-Chebyshev projector, with dependence of their success probability on $\alpha$ given in \Cref{fig:success_prob_alpha}}

{
\subsection{Product of Linear Combinations of Unitaries}}
One may obtain encodings of the non-unitary operators $\hat H_\nu = \frac{\hat{H}-a_{\nu}}{S-a_{\nu}}$ and apply them sequentially.

We note that, for the operators $\hat H_\nu$, the denominator will be absorbed in the sub-normalisation. { Therefore, in practice we are interested in block-encoding the polynomial 
\begin{equation}
    P(\hat H) = \prod_{\nu = 1}^{m}(\hat H - a_\nu)
    \label{eq:prod}
\end{equation}
The polynomial is achieved by succesive applications of $U\left(\hat H-a_{1}\right), \ \ldots,\ U\left(\hat H-a_{m}\right)$, with the projection applied after each $U$ via mid-circuit measurement on the $|0\cdots0\rangle$ state, as shown in \Cref{fig:lcu_projector}.
Given an {($\alpha$,$b$,$0$)} encoding of the Hamiltonian, it is easy to obtain an {($\alpha + |a_\nu|$,$b+1$,$0$)} encoding for each of the terms in eq. \ref{eq:prod} \mbox{\cite{Gilyen2019}}. However, since in most realistic scenarios the Hamiltonian $\hat H$ will include a diagonal term and the values $a_\nu$ are known \textit{a priori} for a given value of $S$, it is more efficient to generate $m$ different ($\alpha_\nu, b, 0$) oracles corresponding to each term in the expansion. This will require one fewer ancilla and improve success probability, as it is always the case that $\alpha_\nu \leq \alpha + |a_\nu|$. This implementation is conceptually simple, requires no additional ancilla beyond those required to block-encode the Hamiltonian and requires ${\Theta}(m)$ gates. The success probability of applying this projector is given by
\begin{equation}
    \begin{split}
        p_\mathrm{succ} &= \left\langle\psi{\left|\prod_{\nu = 1}^m\frac{(\hat H - a_\nu)^2}{\alpha_\nu^2}\right|}\psi\right\rangle\\ 
        &= \sum_i c_i^2 \prod_{\nu = 1}^m\frac{(E_i-a_\nu)^2}{\alpha_\nu^2}.\\
    \end{split}
\end{equation}
In the limit of large $m$, when all excited state contributions are projected out, 
\begin{equation}
    p_\mathrm{succ} \approx c_0^2 \prod_{\nu = 1}^m\frac{(E_0-a_\nu)^2}{\alpha_\nu^2}.
\end{equation}
In general, $E_0 - a_\nu = \mathcal{O}(||\hat H||)$, while using LCU with Pauli terms $\alpha_\nu  = \mathcal{O}(||\hat H||_1)$, which can, in the worst case, be exponentially larger than $||\hat H||$. Therefore, for general Hamiltonians, this block-encoding may have exponentially decaying success probability as the required $m$ for convergence increases.

We note that, much like in the fragmented imaginary time evolution method,\cite{Silva2023} mid-circuit measurement of the ancilla qubits allows the total number of quantum operations to guarantee a successful application of the projector to scale as
\begin{equation}
    Q_m = \sum_{k=1}^{m}\frac{p_\text{succ}(k-1)}{p_\text{succ}(k)}
\end{equation}
where $p_\text{succ}(k)$ is the succes probability of applying the first $k$ polynomial terms successfully, rather than the much larger $1/p_\mathrm{succ}(m)$.

\subsection{Generalised Quantum Singular Value Transform}}
\begin{figure*}
    \includegraphics[width =0.48\textwidth]{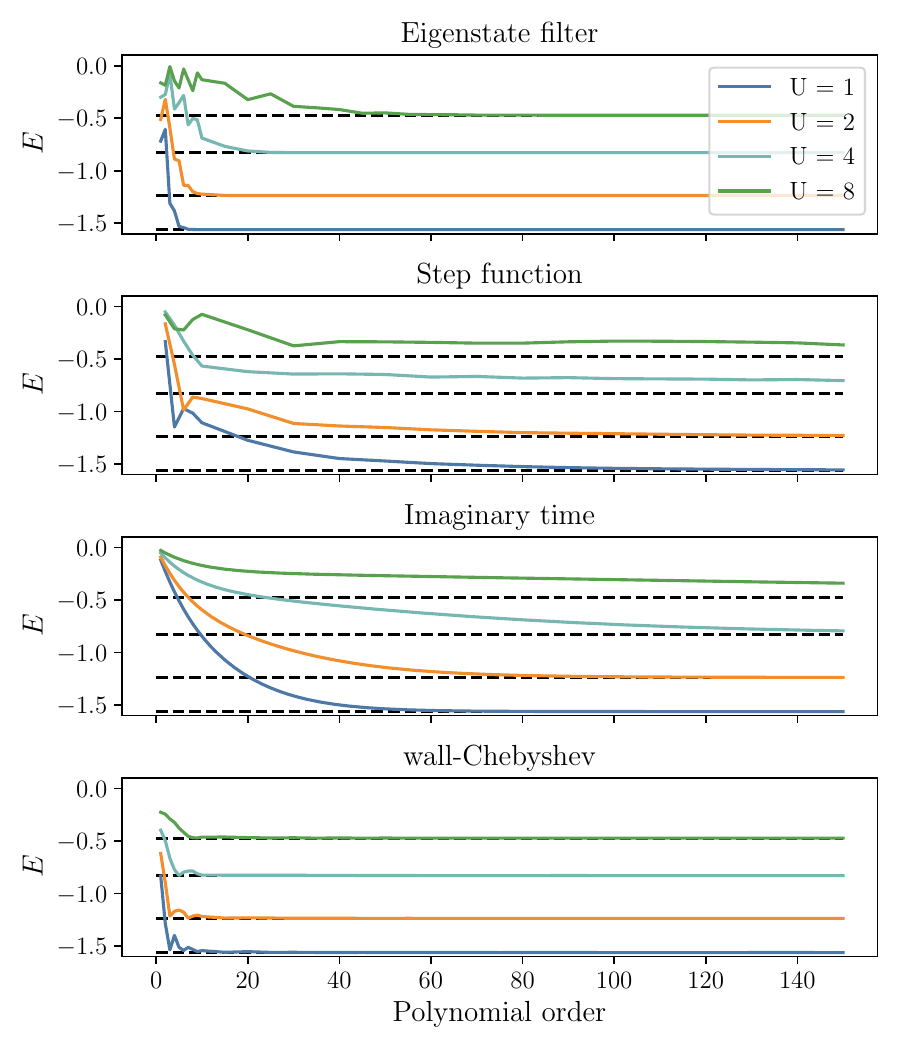}
    \includegraphics[width =0.48\textwidth]{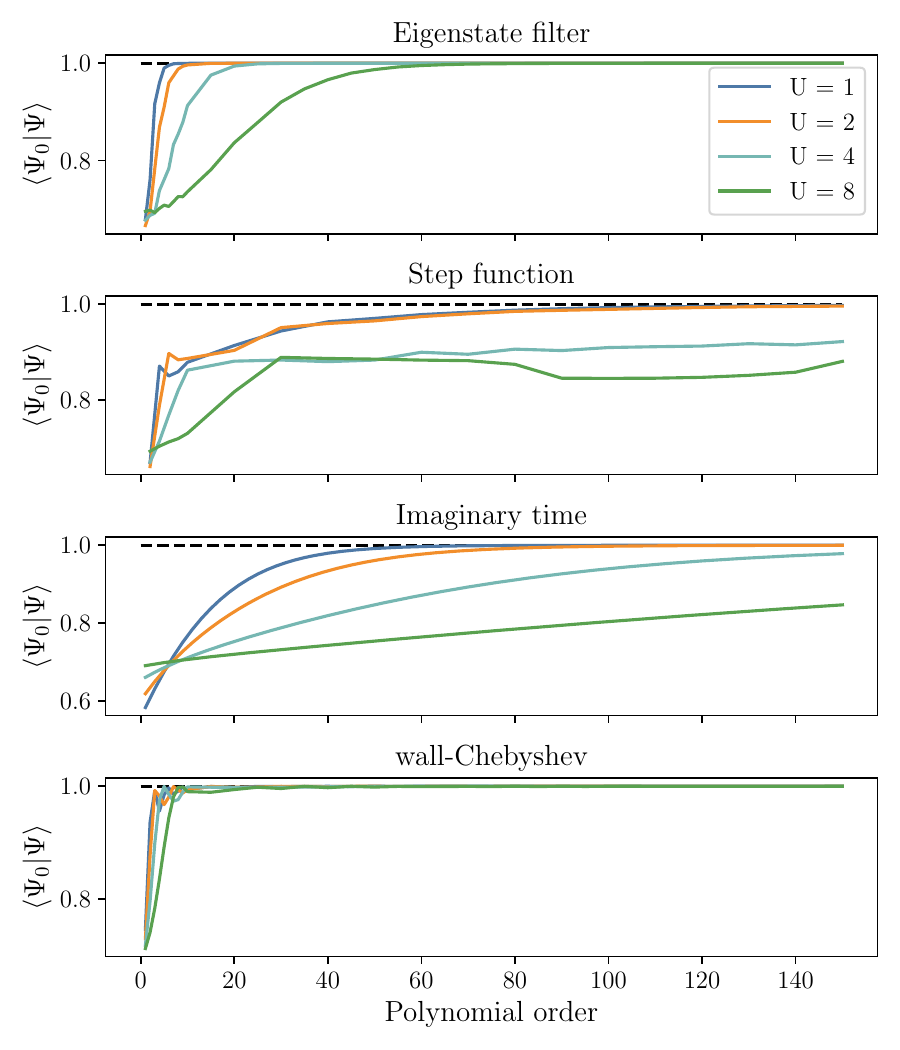}
    \caption{Projector convergence onto the ground state of the two-site Hubbard model with $t = 1$ and different values of $U$.  The left panels show energy convergence, while the right show fidelity with the ground state wavefunction, as a function of polynomial order. In all cases, the ground state energy is known \textit{a priori} and given as a parameter to the projector.}
    \label{fig:hub_2site}
\end{figure*}
A popular approach for block encoding polynomial functions of operators is the Quantum Singular Value Transformation (QSVT),\cite{Gilyen2019} some details of which are given in Appendix~\ref{app:qsvt}. { However, this approach is limited to polynomial functions with well-defined parity.
Therefore, the step function and eigenvalue filter methods can be implemented by QSVT,
but not the imaginary time-evolution and wall-Chebyshev methods, which have both odd and even components. The odd and even parts of the polynomial could be encoded separately, as is customarily done for the (imaginary) time-evolution operator\cite{Martyn2021,Martyn2023,Silva2023}, or the projector could be encoded by generalised Quantum Singular Value Transformation (GQSVT)\cite{Sunderhauf2023, Motlagh2024},  which may be used for arbitrary sums of Chebyshev polynomials, allowing direct implementation of \Cref{eq:small_G}. A short overview of this approach is also given in Appendix~\ref{app:qsvt}.

Like the product approach, encoding an order-$m$ projector with (G)QSVT also requires $m$ calls to the Hamiltonian oracle. However, the success probability is given by
\begin{equation}
    p_\mathrm{succ} = \braket{\psi|P(\hat H/\alpha)^\dagger P(\hat H/\alpha)|\psi},
\end{equation}
which does not in general decay exponentially with $m$, as in the case of product formulas. We note however, that QSVT allows us to encode polynomials of $\hat H/\alpha$, not of $\hat H$ itself. Our target Hamiltonian $\widetilde H$ has $||\widetilde H|| = 1$, but as discussed before block-encoding does not in general achieve a sub-normalisation equal to the spectral norm of the Hamiltonian, so $\alpha > 1$. This shifts the lowest eigenvalue of $\widetilde H /\alpha$ away from -1, where $G_m^\mathrm{wall-Ch}(-1) = 1$, lowering the success probability. We have two options to block-encode the wall-Chebyshev projector in this form.
\begin{figure*}
    \includegraphics[width =0.48\textwidth]{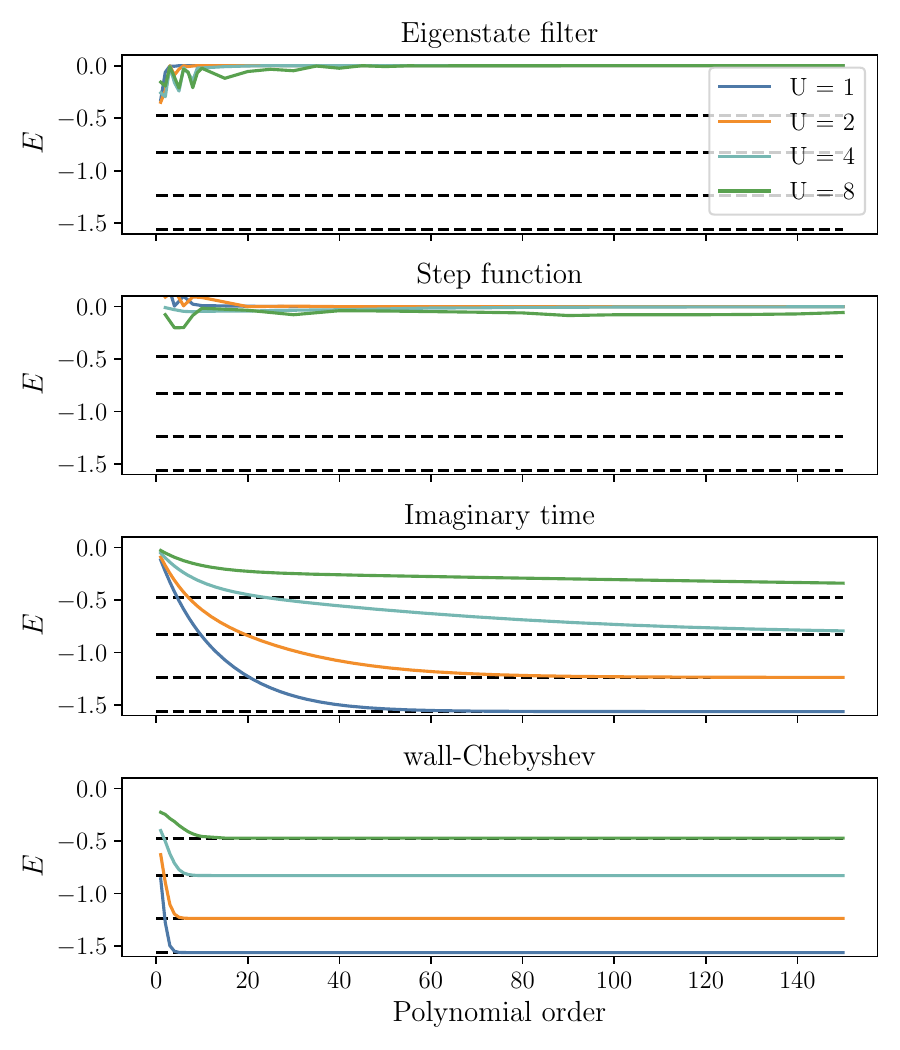}
    \includegraphics[width =0.48\textwidth]{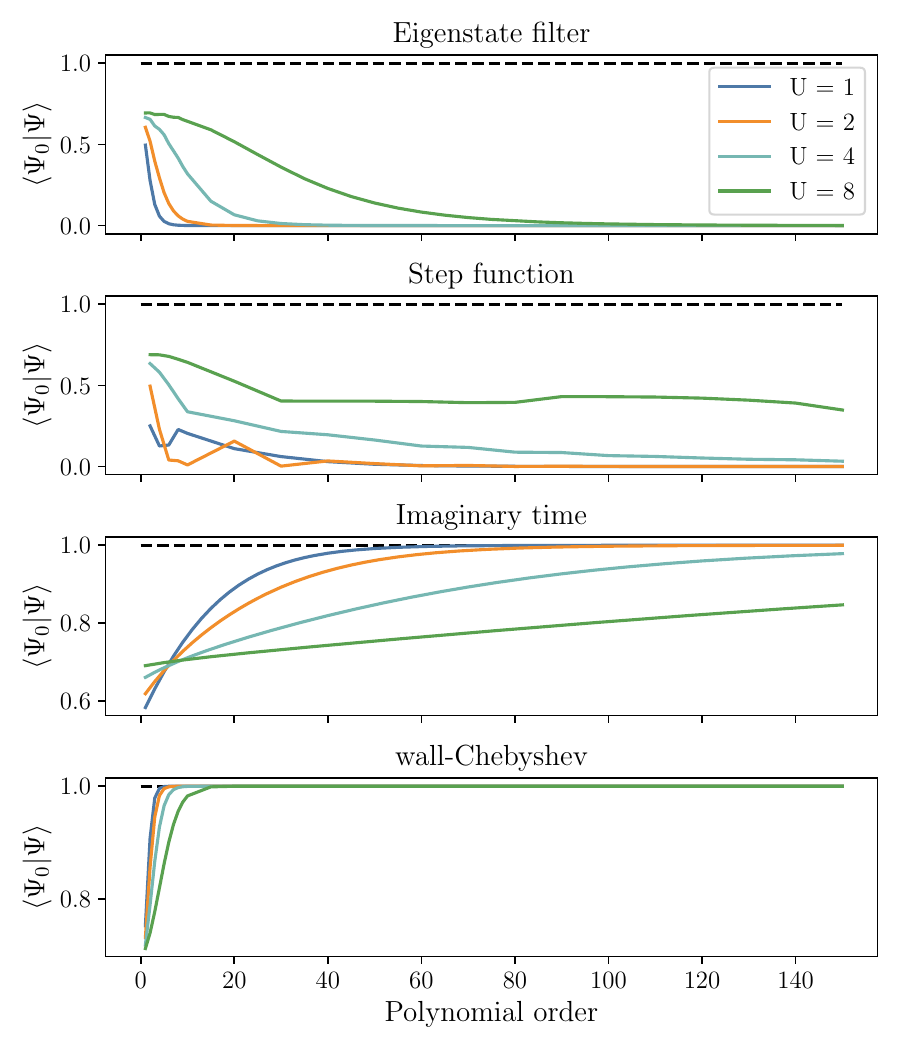}
    \caption{Projector convergence onto the ground state of the two-site Hubbard model with $t = 1$ and different values of $U$. The left panels show energy convergence, while the right show fidelity with the ground state wavefunction, as a function of polynomial order. In all cases, the ground state energy is estimated as the energy of the Hartree--Fock state.}
    \label{fig:hub_2site_unknown}
\end{figure*}

\subsubsection{Direct encoding of the wall-Chebyshev polynomial}
In this approach we block-encode

\begin{equation}
    P_1(\widetilde H/\alpha) = G_m^\mathrm{wall-Ch}\left(\frac{\widetilde H}{\alpha}\right) = \prod_{\nu = 1}^m\frac{\left(\frac{\widetilde H}{\alpha} - \widetilde a_\nu\right)}{(-1 - \widetilde a_\nu)},
\end{equation}
where $\widetilde a_\nu = - \cos\frac{\nu\pi}{m+1/2}$, which will have a success probability 
    \begin{equation}
    \begin{split}
         p_\mathrm{succ} &= \sum_i c_i^2 \left[G_m^\mathrm{wall-Ch}\left(\frac{\widetilde E_i} {\alpha}\right)\right]^2 \\
         &\approx c_0^2 \left[G_m^\mathrm{wall-Ch}\left(-\frac{1} {\alpha}\right)\right]^2
    \end{split}
    \end{equation}
This is the simplest block encoding of the Chebyshev wall-function, but is different from the projector encoded in the product formula discussed above. The value of $G_m^\mathrm{wall-Ch}\left(-\frac{1} {\alpha}\right)$ depends on the relative positioning of $-1/\alpha$ and the nodes in the Chebyshev polynomials and we note from \Cref{fig:success_prob_alpha} that this implementation preserves reasonable success probabilities even for moderately large $m$, unless $-1/\alpha$ is coincidentally close to one of the nodes. However, in general, the larger $\alpha$ is, the farther $-1/\alpha$ is from the ``wall" at $-1$, leading to small values of $G_m^\mathrm{wall-Ch}\left(-\frac{1} {\alpha}\right)$, in particular as $m$ increases. However, this behaviour is non-monotonic, and, corresponds to non-monotonic quality of the projector itself, the practical applicability of which we assess in the upcoming sections. 
\subsubsection{wall-Chebyshev with rescaled nodes}
 In order to encode an equivalent function to the product form, we consider 
 \begin{equation}
P_2(\widetilde H/\alpha) = \prod_{\nu = 1}^m\left(\frac{\widetilde H - \widetilde a_\nu}{\alpha}\right) = \sum_{k = 0}^{m} c_k T_k\left(\frac{\widetilde H}{\alpha}\right).     
 \end{equation} The coefficients $c_k$ may be found numerically. This approach guarantees that the ground state energy is lower than all nodes in the polynomial and has success probability
     \begin{equation}
         p_\mathrm{succ} = \sum_i c_i^2 \left[P_2\left(\frac{\widetilde E_i} {\alpha}\right)\right]^2 \approx c_0^2 \left[P_2\left(-\frac{1} {\alpha}\right)\right]^2
    \end{equation}
 which once again decays rapidly with $\alpha$, limiting the applicability of this projector to general Hamiltonians, although utility may still be found for sparse Hamiltonians, for which $\alpha$ may be brought closer to the spectral norm.}
\begin{figure*}
    \includegraphics[width=\textwidth]{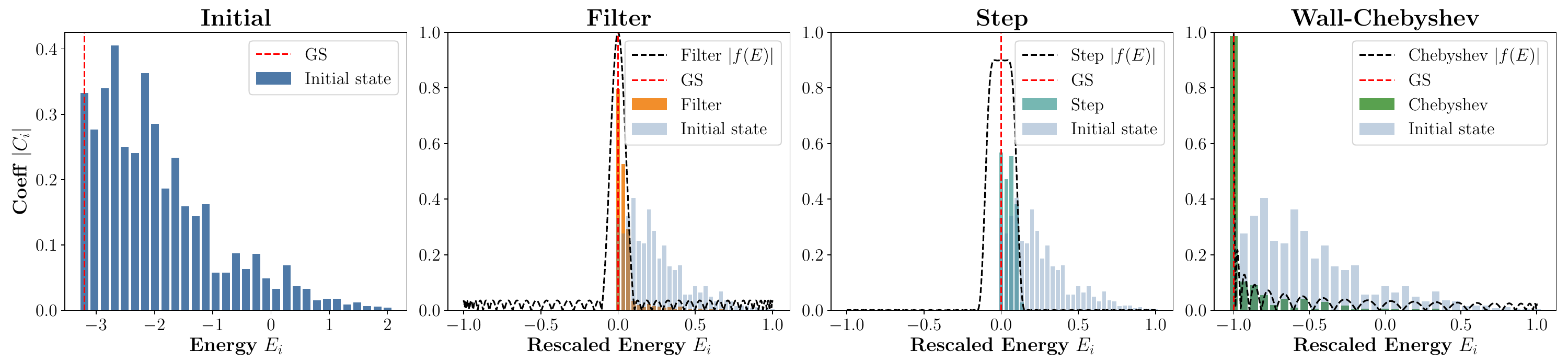}
    \includegraphics[width=\textwidth]{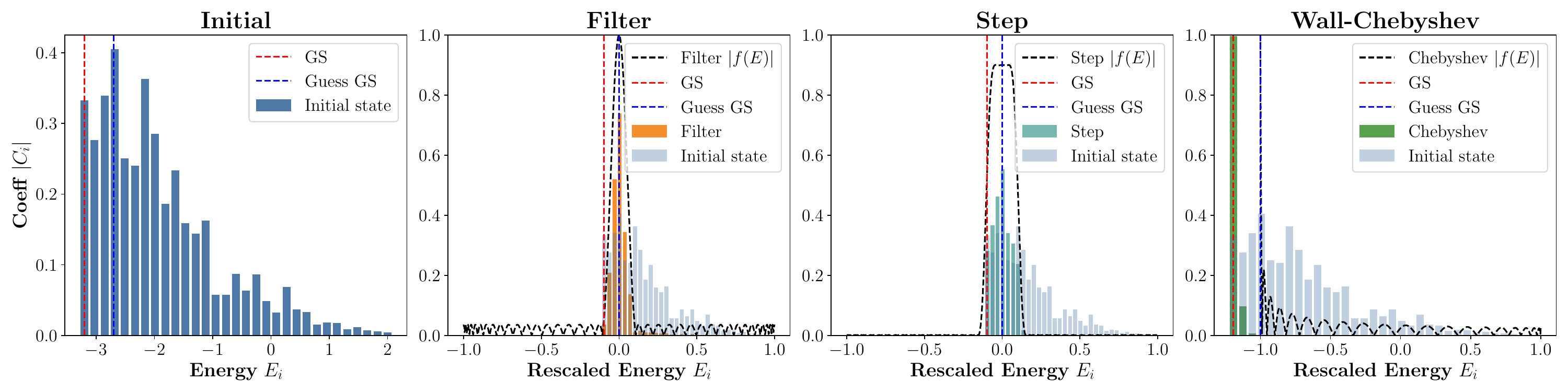}
    \caption{Comparison of different projectors: eigenstate filter (orange), step function (turquoise) and wall-Chebyshev (green) ground state projectors of an arbitrary Hamiltonian $H$, given knowledge of the exact ground state { energy} (top) or an approximation (bottom). The filter and wall-Chebyshev polynomials are approximated by a degree 20 polynomial, while the step function is approximated by a 100 degree polynomial, necessary to capture the flat top of the function. Showing their respective action on an initial state (blue) $|\Psi\rangle = \sum^N_i C_i |\phi_i\rangle$, where each bar corresponds to $|\phi_i\rangle$ and is an eigenvector with an expansion coefficient $C_i$ (shown on the y axis) and a corresponding eigenvalue $E_i$ (shown on the x axis). The projection acts via a renormalisation of $|\Psi'\rangle = \sum_i f(E_i) C_i |\phi_i\rangle$. Where $|\Psi'\rangle$ is the projected state and $|f(E_i)|$ is the absolute value of the respective projection function on the eigenvalues at $E_i$ (dotted black). For each projector, the rescaled initial state is shown for comparison (light blue). The lowest energy in each case is the ground state of $H$.}
    \label{fig:projector_barchart}
\end{figure*}

\subsection{Amplitude Amplification}

{
If the wall-Chebyshev projector is implemented as a product of block encodings with deferred measurement using ancilla registers~\cite{vasconcelos2025methodsreducingancillaoverheadblock} or { GQSVT}, it can be combined with Oblivious Amplitude Amplification (OAA)~\cite{Brassard2002}. OAA boosts the success probability to a constant value using $ q = \Theta\!\left(\frac{1}{\sqrt{p}}\right)$ queries to the projector circuit, where $p$ is the raw success probability. In the case of exponentially decaying success probability, given by $p = \alpha^{-2m}$, this results in $q = \alpha^{m}$ which is still exponential in $m$. Nevertheless, this approach could be practical when applied with $P_1$, which has been shown in certain regimes in Figure~\ref{fig:success_prob_alpha} to achieve a convergent success probability.}

\section{Numerical comparisons}

To assess the practical performance of the wall-Chebyshev projector, we compare it to the step function projector in Ref. \cite{Lin2020b}, the eigenstate filtering approach in Ref. \cite{Lin2020a} and an imaginary time evolution projector based on the Chebyshev expansion in Ref.\ \cite{Silva2023}. Circuit implementations and simulations of the wall-Chebyshev projector incorporating mid-circuit measurements were performed using the wallcheb package~\cite{wallcheb}, within the Guppy quantum circuit framework~\cite{Koch2024}. The resulting circuits were compiled via the multiplexor LCU block-encoding method as described in Ref.~\cite{sze2025hamiltoniandynamicssimulationusing}. The coefficients for the step function projector were obtained using QSPPACK\cite{Dong2021,Wang2022,Dong2024a,Dong2024b}. For imaginary time evolution, there are two free parameters, the polynomial order \textit{per} time-step and the size of the imaginary time-step. To reduce this to one parameter, we use the first-order approximation,
\begin{equation}
e^{-\tau H} \approx I_0(\tau) -2 I_1(\tau) T_1(H),  
\end{equation}
where $I_l$ are modified Bessel functions of the first kind. We set the time-step such that the truncation error of this approximation is less than $0.01$. { For the wall-Chebyshev projector, we consider the projector as encoded by the product formula or $P_2\left(\frac{\widetilde H}{\alpha}\right)$, unless otherwise specified.}

As a first test case, we consider the Hubbard model, with Hamiltonian
\begin{equation}
   \hat{H} = - t\sum_{i,\sigma} (c_{i,\sigma}^\dagger c_{i+1, \sigma} + c_{i+1,\sigma}^\dagger c_{i, \sigma}) + U \sum_i n_{i\uparrow} n_{i\downarrow},
\end{equation}
where $c^\dagger_{i,\sigma}$ and $c_{i,\sigma}$ are creation and annihilation operators for electrons with $m_s = \sigma$ at site $i$, respectively, and $n_{i,\sigma} = c^\dagger_{i,\sigma}c_{i,\sigma}$ are number operators. The { ratio} $U/t$ controls the degree of correlation in the system.

For the two-site Hubbard model at half filling in the site basis with spin projection $m_s = 0$,  we consider first the convergence of the four projectors when the true ground state energy is known, with results shown in \Cref{fig:hub_2site}. We find that in this scenario, the performance of the wall-Chebyshev projector is similar to that of the eigenstate filtering algorithm, with both converging significantly faster than the step function based projector and imaginary time evolution. For large $U$, the wall-Chebyshev projector converges up to six times faster than the eigenstate filtering approach. However, the difference in performance between the four approaches is particularly stark if the true ground-state energy is not known. The lowest energy single-determinant wavefunction in this system, corresponding to the $1^\alpha 2^\beta$ configuration or its spin-flipped pair, has zero energy at any $U/t$ ratio. If we use this as an initial guess for the ground state energy $S$, the performance of the four projectors is shown in \Cref{fig:hub_2site_unknown}. For these results we used a value of $\beta = 1.1$, although testing other values up to $\beta = 1.5$ showed very similar performance. As an example, for U=1, in this mapping $R=3.3$, the true ground state corresponds to $x\approx-1.946$ and the highest excited state to $x\approx 0.552$.

\begin{figure}
\centering
\includegraphics[width=0.8\linewidth]{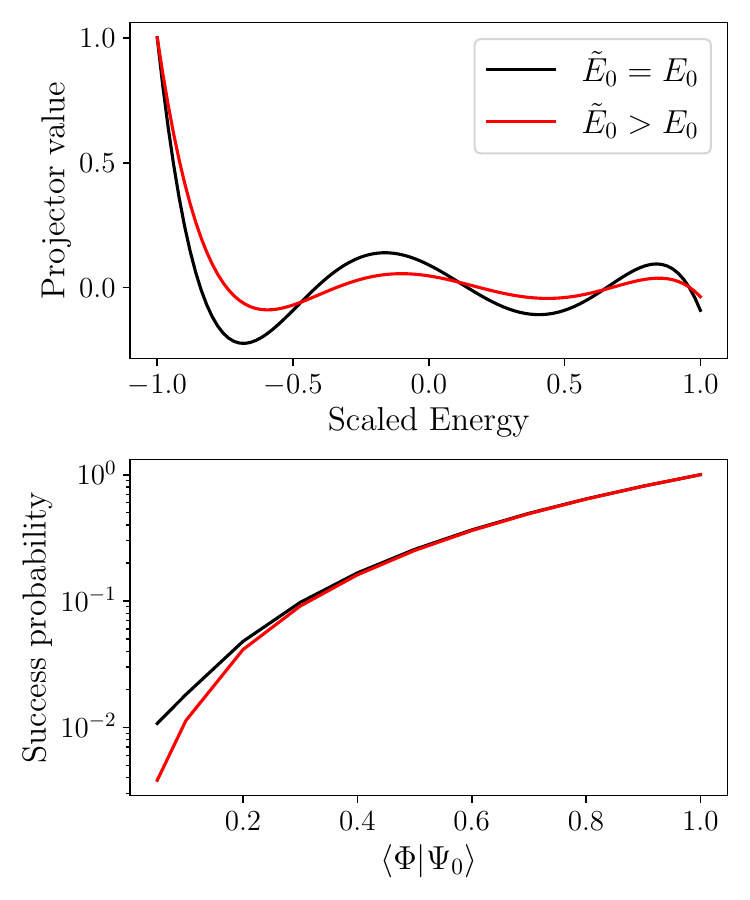}\\
    \caption{Effect of overestimating the ground-state energy in a $m = 5$ quantum wall-Chebyshev projector, given a perfect block encoding with $\alpha = 1$. \textbf{Top:} Value of the projector as a function of scaled energy eigenvalue. { For the red curve, $E^\mathrm{guess}_0$ is 10\% higher than $E_0$.} \textbf{Bottom:} Success probability of applying the projector as a function of initial overlap with the ground state, assuming minimum allowed subnormalisation $\alpha$.}
    \label{fig:success_prob}
\end{figure}
\begin{figure*}
    \includegraphics[width=\textwidth]{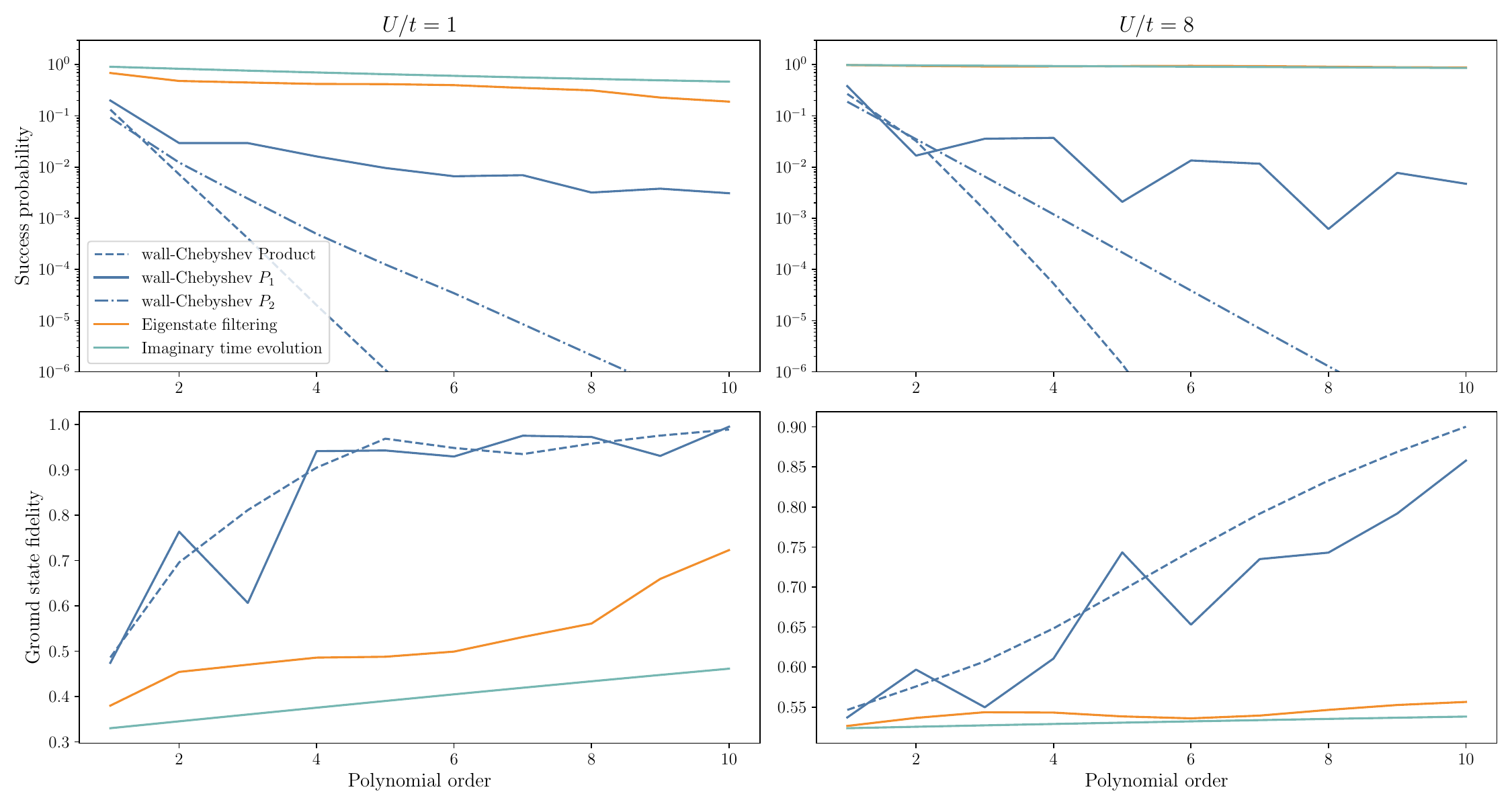}
    \caption{Success probability and ground state fidelity of different projectors{  applied to the 4-site Hubbard lattice with different values of U/t}, including all three encodings of the wall-Chebyshev considered here. In the fidelity plot there is only one line for the wall-Chebishev product encoding, as the fidelity behaviour of the $P_2$ encoding is equivalent.}
    \label{fig:success_prob_real}
\end{figure*}
In this case, the wall-Chebyshev and imaginary time evolution projectors show similar convergence characteristics as before, while both the eigenstate filtering and the step function projector fail to converge to the correct state. We can understand this behaviour by considering the effect of the polynomials onto an arbitrary wavefunction in both scenarios, as seen in \Cref{fig:projector_barchart}. In the case of the eigenstate filter and step function, the guess for the ground state energy is used as the 0 of the rescaled Hamiltonian, while in the wall-Chebyshev case it corresponds to $-1$. If the true ground state is below this value, in the eigenstate filter case it will fall outside the width of the $\delta$ function, causing the contribution of the ground state to be damped. In the step function case, the main issue is the existence of excited states between the ground state and the estimate, which will also be left unchanged by the step function, rather than being damped. In the wall-Chebyshev case, the ground state now falls outside the $[-1,1]$ interval, where the function goes steeply towards infinity. As such, it will continue to be the most undamped contribution to the wavefunction even if the guess energy is not exact. While not shown in \Cref{fig:projector_barchart}, the same argument applies to the conventional imaginary time evolution operator. We note that, due to the rescaling required to apply the Hamiltonian through the LCU, the wall-Chebyshev projector does not in practice over-express the ground state if it lies below $-1$, as could be expected from \Cref{fig:cheby_poly}, but rather more strongly damps other contributions in order to maintain normalisation of the resulting wavefunction. { In effect, the resulting projector is
\begin{equation}
    g'(E) = \frac{g(E)}{g(E_0)}
\end{equation}
}This leads to faster convergence to the ground state, at the cost of lower success probability in applying the operator, as more information is lost upon application. { Consider a normalised wavefunction with only two eigenstate contributions,
\begin{equation}
    \ket{\Psi} = c_0 \ket{\Psi_0} + c_1\ket{\Psi_1}
\end{equation}
If a perfect $\alpha = 1$ block encoding of $||\widetilde H||$ is available, the probability of successfully applying projector $g$ to this wavefunction is given by
\begin{equation}
    p_\mathrm{succ} = |c_0|^2 + |c_1|^2\left|\frac{g(E_1)}{g(E_0)}\right|^2.
\end{equation}
When $E_0^\mathrm{guess} > E_0$, $\left|\frac{g(E_1)}{g(E_0)}\right| < |g(E_1)|$, leading to a reduction in success probability. As can be seen from \Cref{fig:success_prob}, this is not a problem unless the initial overlap with the ground state is quite low.}

{ The previous discussion only considers the number of Hamiltonian applications needed to obtain a certain fidelity with the ground state and does not account for the associated success probability. In \Cref{fig:success_prob_real} we plot the success probability and ground state fidelity of the four projectors considered in \Cref{fig:hub_2site,fig:hub_2site_unknown}, as well as the direct block-encoding of the wall-Chebyshev function, $P_1$, { for the 4-site Hubbard model}. As expected the success probabilities of wall-Chebyshev polynomials decay faster than those of other approaches and are not bounded from below by $c_0^2$, although the $P_1$ implementation retains a reasonable probability of succcess. Although this is not guaranteed by its form, we also find that the $P_1$ approach closely follows the product and $P_2$ approaches in ground state fidelity, suggesting it as a viable alternative. Most importantly, particularly in the strongly correlated $U/t = 8$ case, all other projectors fail to extract the ground state wavefunction for polynomial implementations with $m \leq 10$, making the wall-Chebyshev approach of interest in spite of its decaying success probability. This method therefore introduces a new trade-off space to this problem, where ground states can be recovered with low-depth circuits, at the cost of an increased number of repetitions. In particular in the early-fault-tolerant regime, where quantum computers with very long coherence times may yet be scarce, this is a potentially appealing methodology.}

The Hubbard model is pathological in this scenario, as (a) the single-determinant energy is much higher than the ground state and (b) there are eigenstates of the Hubbard Hamiltonian which are degenerate to this single-determinant, making it impossible for the step function to select a single eigenstate. 

\begin{table}[h!]
    \centering
    \begin{ruledtabular}
    \begin{tabular}{c|c|c|c|c|c|c|c}
    \multirow{2}{*}{Method}&\multirow{2}{*}{$r_\mathrm{H-H}/$\AA}&\multicolumn{3}{c|}{Known $E_0$} & \multicolumn{3}{c}{Unknown $E_0$} \\
    \cline{3-8}
         & & H$_2$ & H$_4$ & H$_6$& H$_2$ & H$_4$ & H$_6$  \\
         \hline
        \multirow{5}{*}{wall-Chebyshev}& 1.00 & 3& 7& 12& 3& 6&8\\
         &1.50 & 5& 11& 22& 3& 7&11\\
         &2.00 & 5& 14& 37& 3& 6&24\\
         &2.50 & 15& 18& 46& 3& 6&12\\
         &3.00 & 10& 25& 49& 3& 6&12\\
         \hline
         \multirow{5}{*}{Eigenstate filter}& 1.00 & 2& 51& -& 2& 53&-\\
         &1.50 & 3& - & -& 44& -&-\\
         &2.00 & 31& - & -& 23& -&-\\
         &2.50 & 17& - & -& 47& -&-\\
         &3.00 & 55& - & -& 27& -&-\\
        \end{tabular}
    \end{ruledtabular}
    \caption{Polynomial order at which the energy of the state obtained by applying a projector first has an error of less than 1 mHartree relative to the true ground state energy for H$_n$ molecules in the STO-3G basis. A maximum polynomial order of 150 was used.}
    \label{tab:hydrogen}
\end{table}

\begin{figure*}
    \includegraphics[width =0.47\textwidth]{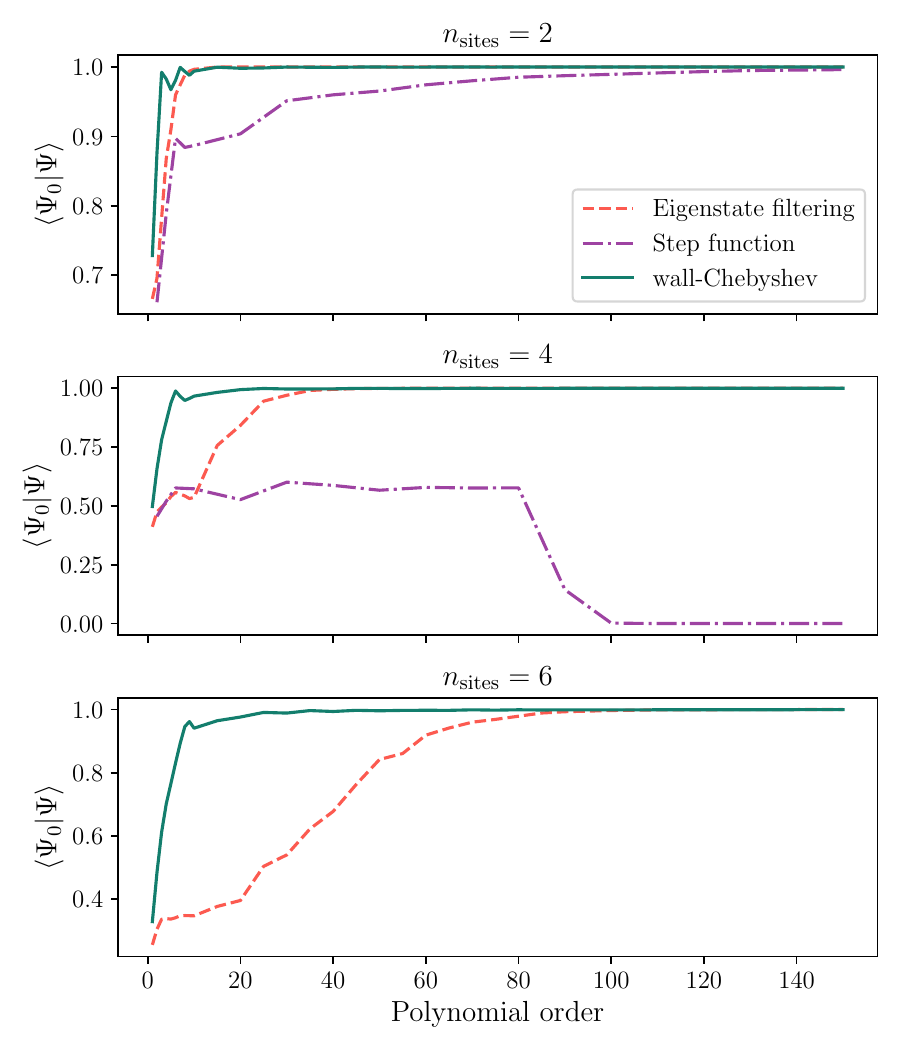}
    \includegraphics[width =0.47\textwidth]{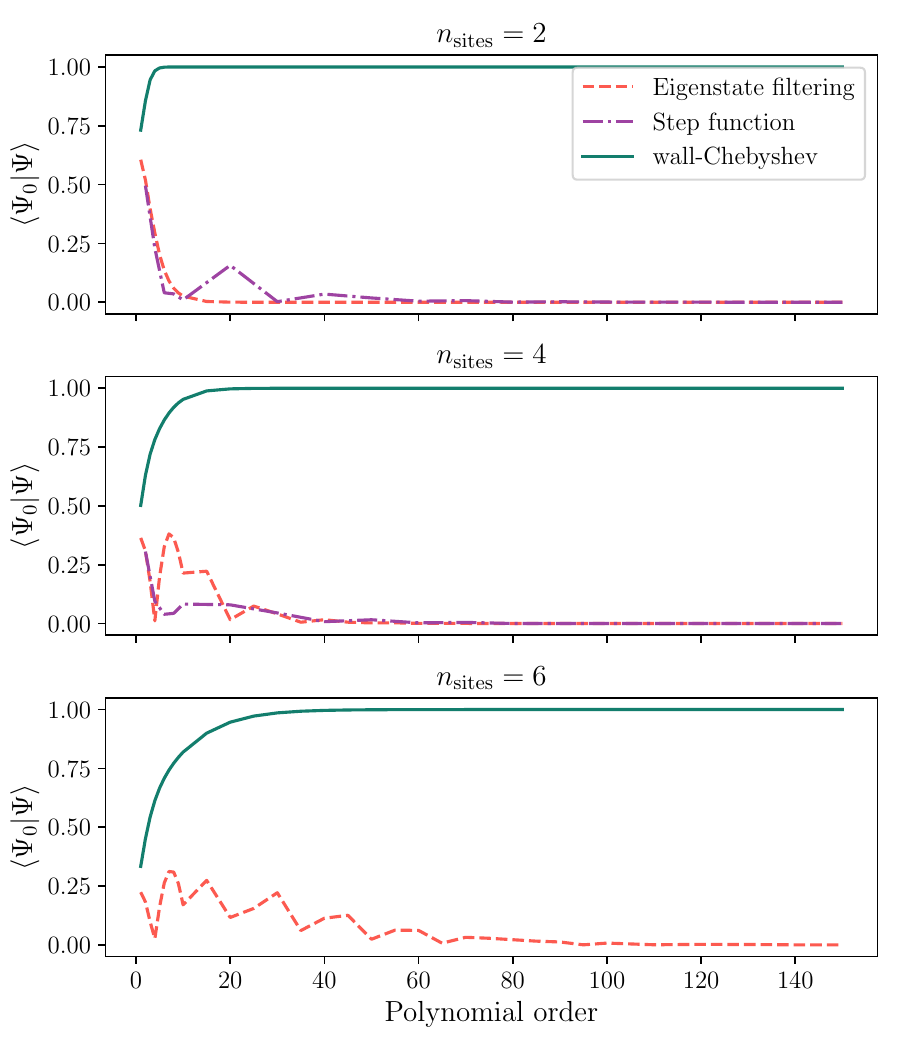}
    \caption{Convergence onto the ground state of the half-filled Hubbard model with $t = 1$, $U=2$, and two (top), four (middle) or six (bottom) sites, using different ground state projectors. The panels show fidelity with the ground state wavefunction, when the ground state energy is known \textit{a priori} (left) or not (right).}
    \label{fig:more_hubbard}
\end{figure*}

In \Cref{fig:more_hubbard}, we consider the performance of the projectors with an increasing number of sites, using the fidelity to the ground state as the convergence metric. We only consider the step function for two- and four-site Hubbard models as the small spacing between the ground and first excited state in larger models made it difficult to converge. As the imaginary time evolution projector is guaranteed to be slower than the wall-Chebyshev, we do not report results for it here. We observe the same patterns as before, with faster convergence of the wall-Chebyshev projector relative to alternative methods as the number of sites increases. Both other algorithms do not converge to the correct state when the HF energy is taken as an approximation of the ground state, and the step function projector additionally fails in the 4-site case even when an exact estimate of the energy is given.
{ The success probability of the projectors follows the trends seen in \Cref{fig:success_prob_real} for all system sizes. In particular, we note that the success probability of the $P_1$ block-encoding is largely system independent, depending most strongly on the strength of correlation in the system.}

As noted above, the Hubbard model is particularly difficult for preexisting methods to get right in the absence of a very good energy estimate. We now consider a system less challenging from this perspective, in the form of hydrogen chains H$_n$ with $2 \leq n \leq 6$ in the STO-3G basis set. In \Cref{tab:hydrogen}, we present the order of each projector needed to obtain errors below 1 mHartree in the ground state energy for each system, restricting ourselves to the more quickly converging eigenstate filter and wall-Chebyshev projectors. At short bond lengths, where Hartree--Fock provides a good approximation for the ground state of the system, the eigenstate filter outperforms the wall-Chebyshev projector for H$_2$ even when the Hartree--Fock energy is used as an estimate for $E_0$. However, as the bond length is stretched and the HF energy moves away from the ground state, the wall-Chebyshev operator becomes once again dominant, requiring a constant polynomial order for convergence at all geometries. Additionally, even if the ground-state energy is known, the eigenstate filter convergence deteriorates more sharply with increasing bond length, likely due to the explicit dependence on the gap between the ground and first excited states. For H$_4$ the difference in performance is much more significant, with the eigenstate filter failing to converge within a polynomial order of 150 for all but the most compressed geometry. For the wall-Chebyshev projector, a maximum order of 50 is sufficient for convergence in all systems.

\section{Conclusion}
We present in this paper a novel ground state projector based on a Chebyshev expansion of the wall function. At finite order, the roots of this expansion have a known closed form, allowing the projector to be expressed as a product of linear Hamiltonian terms. As the roots are known \textit{a priori}, each of these terms can be implemented independently, requiring no additional ancilla relative to the Hamiltonian oracle itself. Like in the case of fragmented imaginary time evolution, this approach allows successful application of each term to be verified, making it possible to abort unsuccessful attempts early. Alternatively, it is possible to encode the projector using GQSVT, which improves the success probability. Decaying success probabilities, however, remain problematic for these ground state projectors, although this can be in principle mitigated through amplitude amplification\cite{Brassard2002} techniques, as is customarily done in QSP-based projectors\cite{Gilyen2019,Martyn2023}.

We show that the wall-Chebyshev projector leads to a polynomial improvement in circuit depth compared to pre-existing approaches based on imaginary time-evolution or alternative function approximations,{ at the cost of potentially exponentially decaying success probability. We find, however, that GQSVT-based implementations maintain non-trivial success probabilities at convergence. Numerically, the wall-Chebyshev projector is shown to outperform projectors based on step functions and delta functions for strongly correlated model systems and a molecular example. Additionally, it preserves a desirable property of imaginary time evolution itself: the ground state amplitude is most weakly damped in the resulting state, regardless of whether a good estimate of the ground state energy is known, allowing the method to converge in the absence of such an estimate.}

\section*{Acknowledgements}

We thank Yuta Kikuchi and Michelle Sze for their helpful comments on this work. MAF gratefully acknowledges Peterhouse, Cambridge for financial support through a Research Fellowship.
\bibliography{bibliography}

\appendix
\onecolumngrid

\section{Gershgorin circle theorem}\label{app:gersh}
The Gershgorin circle theorem\cite{gersgorin_uber_1931} states that, for a complex matrix $A$ with elements $a_{ij}$, one may define the closed Gershgorin discs $D(a_{ii}, R_i) \subseteq \mathbb{C}$, which are centred at $a_{ii}$ and have radii
\begin{equation}
    R_{i} = \sum_{j \neq i} |a_{ij}|.
\end{equation}
Then, every eigenvalue $\lambda_i$ of $A$ must lie in at least one of the Gershgorin disks.

Since the Hamiltonian is a Hermitian matrix, both its eigenvalues and diagonal elements in any basis are always real. The Gershgorin theorem therefore reduces to
\begin{equation}
    \min_{i} (a_{ii} - R_i) \leq \min_{i} \lambda_i \leq \max_i \lambda_i \leq \max_{i} (a_{ii} + R_i) 
\end{equation}
and provides both lower and upper bounds for the Hamiltonian eigenvalues. However, computing these bounds exactly is exponentially expensive and therefore impractical. Instead, we assume that $\max_i(a_{ii} + R_i) = a_{N-1, N-1} + R_{N-1}$ is the extreme of the Gershgorin disk corresponding to the highest excited determinant in the Hilbert space and use this as an estimate of the highest Hamiltonian eigenvalue. This is a likely, but not guaranteed scenario; however, this approximation has been found to work well in classical QMC\cite{zhao2024rapidly}, in particular when combined with a modest stretching factor that mitigates some of the risk of having underestimated the maximal eigenvalue.

\section{Linear Combination of Unitaries}
\label{app:lcu}

\begin{figure}[!htb]
\centering
\includegraphics[width = \textwidth,trim = 2.2cm 24cm 2.2cm 1.8cm, clip]{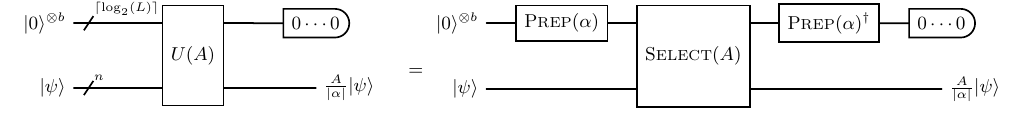}

\caption{LCU Circuit acting on an arbitrary quantum state $|\psi\rangle$ with a rescaled  operator $\frac{A}{|\alpha|}$, when a successful post-selection occurs.}
\label{fig:lcu_figure0}
\end{figure}

Consider an operator $A$, which may be written as
\begin{equation}
    \hat A = \sum_\ell^L \alpha_\ell A_\ell,
\end{equation}
where the operators $A_\ell$ are unitary. An ($\alpha, b, 0$) block-encoding of $A$ can then be implemented using the circuit given in \Cref{fig:lcu_figure0},
where $\alpha = \sum_\ell^L |\alpha_\ell|$ is the 1-norm of the operator $A$ and $b = \lceil \log_2(L) \rceil$.

The LCU block-encoding is made up of two oracles, $\prepare$ and $\select$. The $\prepare$ oracle is a unitary matrix that transforms the all-zero state of the control register to a positive real state $|L\rangle$ that encodes the rescaled coefficient of the LCU. The full unitary matrix of the $\prepare$ oracle is given by:

\begin{equation}
  \prepare = 
  \sum^{L-1}_{\ell = 0} \sqrt{\frac{\alpha_\ell}{|\alpha|}}|\ell \rangle\langle 0 | + \Pi_\perp
  = \begin{bmatrix}
    \sqrt{\frac{\alpha_0}{|\alpha|}} & \cdot & \hdots \\ 
    \sqrt{\frac{\alpha_1}{|\alpha|}} & \cdot & \hdots  \\ 
    \vdots &  \ddots & \hdots   \\ 
    \sqrt{\frac{\alpha_{L-1}}{|\alpha|}} & \cdot  &  \hdots
    \end{bmatrix}.
\,
\label{equation:prepare}
\end{equation}

\begin{figure}[htbp!]
    \includegraphics[width = \textwidth,trim = 2.2cm 22cm 2.2cm 1.8cm, clip]{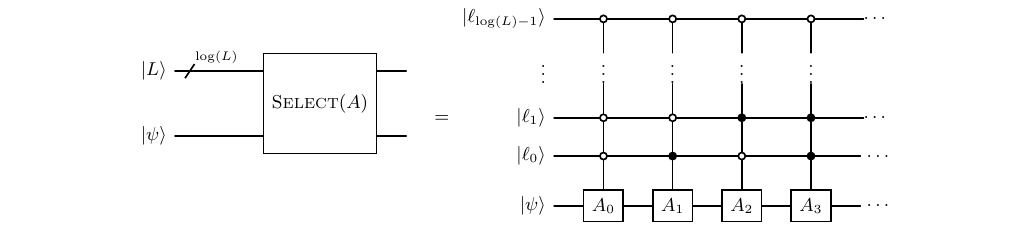}
    \caption{$\select$ circuit.}
    \label{fgr:select}
\end{figure}

It can be seen that we only need to be concerned with the first column of the matrix if acting on the all-zero state.
The $\select$ oracle is a unitary matrix that acts on the state register and the $\prepare$ register. It indexes the unitaries in the LCU onto a control register $|\ell\rangle$ and applies the operator terms to the state register via a multi controlled operation for each term indexed by a unique prepare register bit string:
\begin{equation}
    \select(H):=\sum^{L-1}_{\ell=0} |\ell\rangle \langle \ell | \otimes H_\ell.
\end{equation}
The corresponding circuit is given in \Cref{fgr:select}.

The standard LCU procedure involves initialising the $\prepare$ register in the state $|L\rangle =\sum^{L-1}_{\ell = 0} \sqrt{\frac{\alpha_\ell}{|\alpha|}}|\ell \rangle$. Subsequently, the $\select$ oracle is used to apply the operator terms onto the state register. After this operation, the prepare register is uncomputed, and the state register is measured in the computational basis.

\section{Quantum Singular Value Decomposition}
\label{app:qsvt}

The standard approach to encoding projectors for use in quantum computing is by quantum singular value transformation (QSVT)\cite{Gilyen2019,Martyn2021}. For a given  matrix $A$, with singular value decomposition 
\begin{equation}
    A = \sum_i \lambda_i \ket{\tilde\psi_i}\bra{\psi_i} ,
\end{equation}
we can define a singular value transformation as
\begin{equation}
    f(A) = \begin{cases}
        \sum_i f(\lambda_i) \ket{\tilde\psi_i}\bra{\psi_i},\ \ f \text{ odd}\\
        \sum_i f(\lambda_i) \ket{\psi_i}\bra{\psi_i}.\ \ f \text{ even}\\        
    \end{cases}
\end{equation}
Consider a block-encoding $U$ of $A$, such that
\begin{equation}
    A = \tilde\Pi U \Pi,
\end{equation}
where $\tilde \Pi = \sum_i\ket{\tilde \psi_i}\bra{\tilde \psi_i}$ and $\Pi = \sum_i\ket{\psi_i}\bra{\psi_i}$. The QSVT algorithm uses Quantum Signal Processing (QSP) to encode a polynomial function of $P(A)$ using a phased alternating sequence 
\begin{equation}
    U_{\Phi} \equiv
    \begin{cases} e^{i\phi_1 \tilde\Pi} U\prod_{k=1}^{(d-1)/2} \left( e^{i\phi_{2k} \Pi} U^\dagger e^{i\phi_{2k+1} \tilde\Pi}U \right),\ \ d\ \mathrm{odd}\\
    \prod_{k=1}^{d/2} \left( e^{i\phi_{2k-1} \Pi} U^\dagger e^{i\phi_{2k} \tilde\Pi-I} U\right),\ \ d\ \mathrm{even}
    \end{cases}
\end{equation}
where $\boldsymbol{\Phi} = \{\phi_0,\phi_1,...,\phi_d \} \in \mathbb{R}^{d+1}$ are the phases which encode the polynomial transform. If $U$ is an $(\alpha, b, 0)$-block enncoding of $A$, $U_{\boldsymbol{\Phi}}$ represents a $(1,b,0)-$block encoding of $P^{(SV)}(A/\alpha)$:
\begin{equation}
P^{(SV)}\left( \frac{A}{\alpha} \right) = \tilde \Pi U_{\Phi} \Pi.
\end{equation}
The corresponding phases can be calculated by classical optimisation. For the singular value decomposition to be well-defined, the polynomial $P$ must obey the following constraints:
\begin{itemize}
    \item[(i)] \( P \) has parity \( d \mod 2 \)
    \item[(ii)] \( \forall x \in [-1,1] : |P(x)| \leq 1 \)
    \item[(iii)] \( \forall x \in (-\infty, -1] \cup [1, \infty) : |P(x)| \geq 1 \)
    \item[(iv)] if \( d \) is even, then \( \forall x \in \mathbb{R} : P(ix)P^*(ix) \geq 1 \).
\end{itemize}
{ Polynomials that do not obey the parity requirement can be encoded by a method known as generalised QSVT (GQSVT).\cite{Sunderhauf2023} In order to encode an arbitrary polynomial
\begin{equation}
    P\left(\frac{A}{\alpha}\right) = \sum_{k=0}^{m} c_k T_k\left(\frac{A}{\alpha}\right),
\end{equation}
one must first find the phase factors $\boldsymbol{\Phi}$ needed to encode the polynomial
\begin{equation}
    P'\left(U\right) = \sum_{k=0}^{m} c_k U^k,
\end{equation}
with the same coefficients $c_k$, using generalised quantum signal processing (GQSP)\cite{Motlagh2024}. This can be done efficiently, however it is possible that while $|P(x)| \leq 1$ for all $x \in [-1,1]$, the same is not true for $|P'(x)|$. In this case, the polynomial may need to be further scaled down. Once the phase factors are known a circuit like that shown in fig.~\ref{fig:lcu_projector} can be used to implement $P\left(\frac{A}{\alpha}\right)$. Small changes to the structure are necessary depending on whether the block-encoding is Hermitian or not, but in both cases the same number of total queries to the Hamiltonian block encoding are required.}
\section{Wall-Chebyshev convergence factor}\label{app:gamma}
The convergence factor of a ground state projector is defined as $\gamma = - g'(E_0)$. For the wall-Chebyshev polynomial, we can rewrite \Cref{eq:rescale} for numerical values of the the energy $E$ as 
\begin{equation}
    x = 2\frac{E - E_0}{R} - 1.
\end{equation}
Therefore,
\newcommand\at[2]{\left.#1\right|_{#2}}
\begin{align}
   \at{\frac{d g_m^\mathrm{wall-Ch}(E)}{dE}}{E=E_0} = \frac{2}{R}\at{\frac{d G_m^\mathrm{wall-Ch}(x)}{dx}}{x=-1},
\end{align}
with $G_m^\mathrm{wall-Ch}$ and $g_m^\mathrm{wall-Ch}$ defined according to \Cref{eq:small_G,eq:big_G}, respectively. The derivatives of Chebyshev polynomials are given by
\begin{align}
        &\frac{dT_0(x)}{dx} = 0, \label{eq:derivs1}\\
        &\frac{dT_1(x)}{dx} = 1, \label{eq:derivs2}\\
        &\frac{dT_{2k}(x)}{dx} = (2k) \cdot 2 \sum_{j=1}^k T_{2j-1}(x),\ \ \forall k \in \mathbb{N}_{+}, \label{eq:derivs3}\\
        &\frac{dT_{2k+1}(x)}{dx} = (2k+1) \left[T_0(x) + 2\sum_{j=1}^k T_{2j}(x)\right],\ \ \forall k \in \mathbb{N}_{+}.  \label{eq:derivs4} 
\end{align}
The derivative of the wall function projector is given by
\begin{align}
    \at{\frac{d G_m^\mathrm{wall-Ch}(x)}{dx}}{x=-1} = \frac{1}{2m+1}\left[\sum_{k=0}^{m} (2-\delta_{k0}) (-1)^k \at{\frac{dT_k(x)}{dx}}{x=-1}\right].
\end{align}
Substituting \Cref{eq:derivs1,eq:derivs2,eq:derivs3,eq:derivs4} and considering the case where $m=2n+1$ for simplicity, we find that
\begin{align}
\begin{split}
    \at{\frac{d G_m^\mathrm{wall-Ch}(x)}{dx}}{x=-1} &= \frac{1}{2m+1}\left\{\sum_{k=1}^{n} \left[2 (-1)^{(2k)} (2k)\cdot 2 \sum_{j=1}^k T_{2j-1}(-1)\right]\right.\\
    &+ \left.\sum_{k=1}^{n} \left[2 (-1)^{(2k+1)} (2k+1)\left(T_0(-1) + 2 \sum_{j=1}^k T_{2j}(-1)\right)\right]-2\right\}\\
    &=\frac{1}{2m+1}\left\{\sum_{k=1}^{n} \left[2 (-1)^{(2k)} (2k)(-2k)\right]\right.\\
    &+ \left.\sum_{k=1}^{n} \left[2 (-1)^{(2k+1)} (2k+1)(2k+1)\right]-2\right\}\\
    &= -\frac{2}{2m+1}\left[{1} + \sum_{k=1}^{n}(2k)^2 + \sum_{k=1}^{n}(2k+1)^2\right]\\
    &=-\frac{2}{2m+1}\sum_{k=0}^m k^2\\
    &=-\frac{2}{2m+1}\frac{m(m+1)(2m+1)}{6}\\
    &=-\frac{m(m+1)}{3}.
\end{split}
\end{align}
The convergence factor is therefore
\begin{equation}
    \gamma = - \frac{2}{R}\left[-\frac{m(m+1)}{3}\right]= \frac{2m(m+1)}{3R}.
\end{equation}
An equivalent calculation may be carried out for even $m$, leading to the same value of $\gamma$.
\section{Complexity Analysis of Other Projectors}
\label{app:complexity}

\subsection{Imaginary time evolution}
We follow the procedure of Ref. \cite{Silva2023} to derive the scaling of a QSP approximation to the imaginary time evolution operator, for Hamilonians with $||H|| \leq 1$. Given a Hamiltonian with $||H|| > 1$, it is possible to rescale it to\cite{Silva2023}

\begin{equation}
    H' = \frac{H - \bar \lambda I}{\Delta \lambda},
\end{equation}
where $\bar \lambda = (\lambda^+ + \lambda^-)/2$, $\Delta \lambda = \lambda^+ - \lambda ^-$ and $\lambda^+ (\lambda ^-)$ is an upper (lower) bound to the largest (smallest) eigenvalue of $H$. The transformation $e^{-\tau H}$ can be recovered by $e^{-\tau' H'}$ where $\tau' = \Delta \lambda \tau$.

The imaginary time evolution operator can be expressed in terms of Chebyshev polynomials via the Jacobi--Anger expansion, as
\begin{equation}
    f (H) = e^{-\tau (H - E_0)} = I_0(\tau) + 2 \sum_{k=1}^{\infty} (-1)^k I_k(\tau) T_k(H-E_0),
\end{equation}
where $I_k$ are the modified Bessel functions of the first kind. The error $\epsilon_{tr}$ of approximately synthesising $f(\hat H)$ through a Chebyshev expansion \mbox{$\tilde{f}_q(H) = \sum_{k=0}^{q} b_k T_k(H)$} is given by\cite{ELLIOTT198749,Silva2023}
\begin{equation}
    \epsilon = \max_{-1 \leq \lambda \leq 1} |f(\lambda) - \tilde{f}(\lambda)| = \frac{\left| f^{(q+1)}(\xi) \right|}{2^{q} (q + 1)!}
\end{equation}
for some $\xi \in (-1,1)$. Therefore,
\begin{equation}
    \epsilon_{tr} \leq \frac{\max_{\lambda \in [\lambda_{\min}, \lambda_{\max}]} |f^{(q+1)}(\lambda)|}{2^{ q}(q+1)!}. 
\end{equation}
where $f^{(q+1)}$ is the $(q+1)-$th derivative of $f$. 
Given
\begin{equation}
| f^{(q+1)}(\lambda) | = \tau^{(q+1)} e^{-\tau \lambda} \leq  \tau^{(q+1)},
\end{equation}
\begin{equation}
    \epsilon_{tr} \leq \frac{\tau^{q+1}}{2^{q}(q + 1)!} \leq \sqrt{\frac{2}{\pi(q + 1)}} \left( \frac{e\tau}{2(q + 1)} \right)^{q+1} \leq \left( \frac{\tau e}{2q} \right)^{q},
\end{equation}
where we have used Stirlings approximation to the factorial and assumed $e\tau/2 \leq q$. Solving this equation for $q$ leads to the query complexity of the QSP encoding of a polynomial of order $q$ approximation of the imaginary time evolution operator.

For $t \in \mathbb{R}$ and $\varepsilon \in (0,1)$, let us define the number $ r(t, \varepsilon) \geq t $ as the solution to the equation 

\begin{equation}
\varepsilon = \left(\frac{t}{r}\right)^r : r \in (t, \infty).
\end{equation}

It can be shown that\cite{Gilyen2019}
\begin{equation}
    \begin{split}
        &\forall t \geq \frac{\ln(1/\varepsilon)}{e}, \ \ r(t,\varepsilon) \leq et,\\
        &\forall t \leq \frac{\ln(1/\varepsilon)}{e},\ \ r(t,\varepsilon) \leq \frac{4\ln(1/\varepsilon)}{\ln (e + \ln(1/\varepsilon)t}
    \end{split}
\end{equation}
and therefore
\begin{equation}
    r(t, \varepsilon) = \mathcal{O}\left(t + \frac{\ln(1/\varepsilon)}{\ln (e + \ln(1/\varepsilon)t}\right).
\end{equation}

In this particular case, $t = \tau e /2$ and therefore
\begin{equation}
    q(\tau, \epsilon_\mathrm{tr}) = \mathcal{O}\left( \frac{e \tau}{2} + \frac{\ln(1/\epsilon_\mathrm{tr})}{\ln(e + 2\ln(1/\epsilon_\mathrm{tr})/(e\tau))} \right).
\end{equation}

If the aim of ITE is to obtain the ground state of the system, then the propagation time $\tau$ depends on the desired fidelity $f(\tau)$. By the same reasoning as before,
\begin{equation}
    \begin{split}
        \epsilon(\tau) &= 1 - f(\tau) \\
        &= 1- \frac{1}{\sqrt{1+\sum_{i>0} \left(\frac{c_i}{c_0}\right)^2e^{-2\tau(E_i - E_0)}}}\\
        &\leq \frac{1}{c_0}\frac{{\sqrt{\sum_{i>0} c_i^2e^{-2\tau(E_i - E_0)}}}}{{1+\sqrt{\sum_{i>0} \left(\frac{c_i}{c_0}\right)^2e^{-2\tau(E_i - E_0)}}}}.
    \end{split}
\end{equation}
Therefore, 
\begin{equation}
    \tau = \mathcal{O} \left(\Delta^{-1} \log (\epsilon^{-1} c_0^{-1})\right).
\end{equation}

Assuming this corresponds to the regime where $\tau e/2 > \log(1/\epsilon_\mathrm{tr})/e$, this implies that
\begin{equation}
    q(\epsilon) = \mathcal{O}\left(\Delta^{-1} \log (\epsilon^{-1} c_0^{-1})\right).
\end{equation}
{If the ground state is known exactly, and $\hat H-E_0$ is block-encoded with sub-normalisation $\alpha$, the success probability is given by
\begin{equation}
    p_\mathrm{succ} = \sum_i c_i^2 \exp\left(-\frac{\tau(E_i - E_0)}{\alpha}\right) \geq c_0^2
\end{equation}
However, if only a lower bound on $E_0$ is known, $E_0 - \Delta_0$, then
\begin{equation}
    p_\mathrm{succ} = \sum_i c_i^2 \exp\left(-\frac{\tau(E_i - E_0+\Delta_0)}{\alpha}\right) \geq c_0^2\exp\left(-\frac{\tau\Delta_0}{\alpha}\right)
\end{equation}}
\subsection{Eigenstate filtering}

The eigenstate filtering algorithm of Lin and Tong\cite{Lin2020a} uses the polynomial of order $2l$ given in \Cref{eq:R_l} to approximate the projector onto the eigenstate with eigenvalue $\lambda$:
\begin{equation}
    P_\lambda \approx R_l (\widetilde H; \widetilde \Delta_\lambda),
\end{equation}
where $\widetilde H = (H - \lambda I)/(\alpha + |\lambda|)$ and $\widetilde \Delta_\lambda$ is the separation of the $\lambda$ eigenvalue from the rest of the spectrum of $\widetilde H$. The $R_l$ polynomial has the property (Ref. \onlinecite{Lin2020a}, Lemma 2) that
\begin{equation}
    \begin{split}
        |R_l(x;\Delta)| \leq 2e^{-\sqrt{2}l\Delta}, \ \ &\forall\ 0 < \Delta \leq \frac{1}{12}\ \mathrm{and}\\
        &\forall x \in [-1, -\Delta] \cap [\Delta, 1]. 
    \end{split}
\end{equation}
This in turn guarantees that
\begin{equation}
    |R_l(\widetilde H,\widetilde \Delta_\lambda) - P_\lambda| \leq 2e^{-\sqrt{2}l\widetilde \Delta},
\end{equation}
so in order to obtain the ground state with fidelity higher than $1-\epsilon$ one requires a number of queries to the Hamiltonian oracle that scales as
\begin{equation}
    q(\epsilon) \propto l = \mathcal{O}\left(\Delta^{-1} \log(\epsilon^{-1} c_0^{-1})\right).
\end{equation}
This is the same asymptotic scaling as the ITE projector. { The success probability of applying this projector is bounded from below by ${c_0^2}$.\cite{Lin2020a}}

\subsection{Step function projector}

The construction of the step function projector in \cite{Lin2020b} is based on constructing a block encoding of the reflection operator,
\begin{equation}
    R_{< \mu} = \sum_{k; \lambda_k<\mu} \ket{\Psi_k}\bra{\Psi_k} - \sum_{k; \lambda_k>\mu} \ket{\Psi_k}\bra{\Psi_k}.
\end{equation}

Given an ($\alpha$, $m$, 0)-block-encoding of the Hamiltonian $H$, one can easily construct an (($\alpha+|\mu|$, $m+1$, 0)-block-encoding of $H-\mu I$ and then, using QSP, an (1, $m+2$, 0)-block encoding of the polynomial $-S(\frac{H-\mu I}{\alpha+|\mu|};\delta, \epsilon)$. This is a polynomial of order $l = \mathcal{O}\left(\delta^{-1} \log(\epsilon^{-1})\right)$, which is defined such that
\begin{align}
&\forall x \in [-1,1],\  |S(x;\delta,\epsilon) \leq 1,\\
&\forall x \in [-1, -\delta] \cup [\delta,1],\  |S(x;\delta,\epsilon) - \mathrm{sgn}(x)| \leq \epsilon.
\end{align}
When the separation between $\mu$ and any eigenvalue is at least $\Delta/2$ and $\delta = \Delta/4\alpha$, this corresponds to a (1,$m+2$, $\epsilon$)-block-encoding of $R_{<\mu}$.

An (1, $m+3$, $\epsilon/2$) encoding of the projector
\begin{equation}
     P_{< \mu} = \sum_{k; \lambda_k<\mu} \ket{\Psi_k}\bra{\Psi_k} = \frac{1}{2}(R_{<\mu} + I)
\end{equation}
can then be block-encoded using the encoding of $R_{<\mu}$ and an additional ancilla. Given the scaling of the polynomial $S$ and the definition of $\delta$, each of these encodings requires $\mathcal{O}(\frac{\alpha}{\Delta}\log(\epsilon^{-1}))$ calls to the block encoding of the Hamiltonian. Given an initial wavefunction with overlap $\braket{\Psi_0|\Phi} \geq c_0$ with the ground state, for final fidelity $1-\epsilon$, an error less than $c_0\epsilon/2$ is needed in the projector.

\begin{equation}
    \frac{\braket{\Psi_0|\widetilde P_{< \mu}|\Phi}}{||\widetilde P_{< \mu}\ket{\Phi}||} \geq \frac{|\braket{\Psi_0|\Phi}| - c_0 \epsilon/2}{|\braket{\Psi_0|\Phi}| + c_0\epsilon/2}  \geq 1 - \frac{c_0\epsilon}{|\braket{\Psi_0|\Phi}| } \geq 1-\epsilon.
\end{equation}
Therefore, the number of calls to the Hamiltonian oracle overall scales as $\mathcal{O}\left(\alpha\Delta^{-1} \log(c_0^{-1}\epsilon^{-1})\right)$. The success probability is given by
\begin{equation}
    ||\widetilde P_{< \mu}\ket{\Phi}||^2 \geq c_0^2 \left(1-\frac{\epsilon}{2}\right)^2.
\end{equation}

\end{document}